\documentclass[aps,pra,groupedaddress,tightenlines]{revtex4}

\usepackage{amsmath}
\usepackage[english]{babel}

\begin{document}

\title{Scheme for fault-tolerant holonomic computation on stabilizer codes}

\author{Ognyan Oreshkov$^{(1,4)}$, Todd A. Brun$^{(1,2)}$, Daniel A.
Lidar$^{(1,2,3)}$}

\affiliation{$^{(1)}$Department of Physics, $^{(2)}$Communication
Science Institute, $^{(3)}$Department of Chemistry,\\
Center for Quantum Information Science \& Technology,
\\University of Southern California, Los Angeles, California
90089, USA\\
$^{(4)}$Grup de F\'{i}sica Te\`{o}rica, Universitat Aut\`{o}noma de
Barcelona, 08193 Bellaterra (Barcelona), Spain}
\date{\today}
\begin{abstract}

This paper generalizes and expands upon the work \cite{OBL08} where
we introduced a scheme for fault-tolerant holonomic quantum
computation (HQC) on stabilizer codes. HQC is an all-geometric
strategy based on non-Abelian adiabatic holonomies, which is known
to be robust against various types of errors in the control
parameters. The scheme we present shows that HQC is a scalable
method of computation and opens the possibility for combining the
benefits of error correction with the inherent resilience of the
holonomic approach. We show that with the Bacon-Shor code the scheme
can be implemented using Hamiltonian operators of weight 2 and 3.

\end{abstract}

\maketitle


\section{Introduction}

There are two main sources of errors in quantum
computers---environment-induced decoherence and imperfect control.
It has been shown that if the errors of each type are sufficiently
uncorrelated and their rates are below a certain threshold, it is
possible to implement reliably an arbitrarily long computational
task with a modest resource overhead \cite{Shor96, DVS96, ABO98,
Kit97, KLZ98, Got97', Got97, Pre99}. This result, known as the
quantum accuracy threshold theorem, is based on the idea of quantum
error correction (QEC) \cite{Shor95, Steane96}---a universal
software strategy to combat noise in quantum computers.

In addition to the software approach, there have also been proposals
to deal with the effects of noise by hardware methods that provide
robustness through their inherent properties. One such method is
holonomic quantum computation (HQC)\cite{ZR99, PZR99}---an
adiabatic, all-geometric method of computation that uses non-Abelian
generalizations \cite{WZ84} of the Berry phase \cite{Berry84}. It
has been shown that due to its geometric nature, this approach is
robust against various types of errors in the control parameters
driving the evolution \cite{CGSV03, CP03, SZZ04, FGL05, ZZ05}, and
thus provides a degree of built-in resilience at the hardware level.

In Ref.~\cite{WZL05} HQC was combined with the method of
decoherence-free subspaces (DFSs) \cite{Dua98, ZR97, LCW98}, which
was the first step towards systematic error protection in
conjunction with the holonomic approach. DFSs provide
\textit{passive} protection, in the sense that no syndrome
measurement and feedback are required. The DFS approach is applicable
provided there is a symmetry in the system-bath coupling, which
results in a sufficiently large decoupled subspace in which quantum
information can be stored. Collective decoherence is a well-known
example of such a symmetry (for a review see
Ref~\cite{LidarWhaley:03}). In the absence of symmetries, the
addition of syndrome measurements and feedback is useful and leads
to the approach based on \textit{active} error correction
\cite{Shor95, Steane96}. Active error correction is also the basis
of quantum fault tolerance, which is essential for the scalability
of any method of computation. Even if we assume that the system is
perfectly protected from environment-induced errors, when the size
of the circuit increases, errors due to imperfect operations would
accumulate detrimentally unless they were corrected. Therefore,
scalability of HQC requires combining the holonomic approach with
active error correction.

In Ref.~\cite{OBL08}, we reported a scheme which combines HQC with
the techniques for fault-tolerant computation on stabilizer codes.
This demonstrated for the first time that HQC is a scalable method
of computation. The scheme uses Hamiltonians which are elements of
the stabilizer---or in the case of subsystem codes, elements of the
stabilizer and gauge groups. Encoded gates are implemented by slowly
varying the Hamiltonians along suitable paths in parameter space,
such that from the point of view of the basis of the full Hilbert
space, the states in each eigenspace undergo the same transversal
transformation. On certain codes such as the 9-qubit Shor code
\cite{Shor95} or its subsystem generalizations \cite{Bac06, BC06},
universal computation according to our scheme can be implemented
with Hamiltonians of weight 2 and 3.

This paper generalizes and expands upon the work \cite{OBL08}. We
provide details on various proofs sketched in Ref.~\cite{OBL08},
clarify many points and analyze properties of the scheme that were
not discussed there. We examine in detail the construction for the
Bacon-Shor code \cite{Bac06, BC06}, discuss the adiabatic
approximation for different parametrizations of the Hamiltonians,
and provide explicit calculations of the holonomy in the
implementation of the $Z$ (phase-flip) gate for two different types
of interpolation.

\section{Preliminaries}

\subsection{Holonomic quantum computation}\label{SecHQC}

Let $\{H_{\lambda}\}$ be a family of Hamiltonians on an
$N$-dimensional Hilbert space, which is continuously parametrized by
a point $\lambda$ in a control-parameter manifold $\mathcal{M}$.
Assume that the family has the same degeneracy structure, i.e.,
there are no level crossings. The Hamiltonians can then be written
as $H_{\lambda}=\sum_{n=1}^{R}\varepsilon_n(\lambda)\Pi_n(\lambda)$,
where $\{\varepsilon_n(\lambda)\}_{n=1}^{R}$ are the $R$ different
$d_n$-fold degenerate eigenvalues of $H_\lambda$, ($\sum_{n=1}^{R}
d_n=N$), and $\Pi_n(\lambda)$ are the projectors on the
corresponding eigenspaces. If the parameter $\lambda$ is changed
adiabatically, a state which initially belongs to an eigenspace of
the Hamiltonian will remain in the corresponding eigenspace as the
Hamiltonian evolves. The unitary evolution that results from the
action of the Hamiltonian $H(t):=H_{\lambda(t)}$ is
\begin{gather}
U(t)=\mathcal{T}\textrm{exp}(-i\int_0^t d\tau H(\tau)) =
\oplus_{n=1}^R e^{i\omega_n(t)}U^{\lambda}_{A_n}(t),
\label{adiabaticevolution}
\end{gather}
where $\omega_n(t)=-\int_0^td\tau\varepsilon_n(\lambda(\tau))$ is a
dynamical phase, and the intrinsically geometric operators
$U^{\lambda}_{A_n}(t)$ are given by the following path-ordered
exponents:
\begin{equation}
U^{\lambda}_{A_n}(t)=\mathcal{P}\textrm{exp}(\int_{\lambda(0)}^{\lambda(t)}A_n).\label{openpathholonomy}
\end{equation}
Here $A_n$ is the Wilczek-Zee connection \cite{WZ84} for the $n$-th
eigenspace, $A_n=\sum_\mu A_{n,\mu} d\lambda^\mu$, where $A_{n,\mu}$
has matrix elements \cite{WZ84}
\begin{equation}
(A_{n,\mu})_{\alpha\beta}=\langle n\alpha;
\lambda|\frac{\partial}{\partial
\lambda^\mu}|n\beta;\lambda\rangle.\label{matrixelementsA}
\end{equation}
The parameters $\lambda^\mu$ are local coordinates on $\mathcal{M}$
($1\leq\mu\leq \textrm{dim}\mathcal{M}$) and $\{|n\alpha;
\lambda\rangle\}_{\alpha=1}^{d_n}$ is an orthonormal basis of the
$n$-th eigenspace of the Hamiltonian at the point $\lambda$.

When the path $\lambda(t)$ forms a loop $\gamma(t)$,
$\gamma(0)=\gamma(T)= \lambda_0$, the unitary matrix with components
$(U_n^{\gamma})_{\alpha\beta}$ appearing in
\begin{equation}
\sum_{\alpha\beta}(U_n^{\gamma})_{\alpha\beta}|n\alpha;0\rangle\langle
n\beta;0|\equiv
U^{\lambda}_{A_n}(T)=\mathcal{P}\textrm{exp}(\oint_{\gamma}A_n)\label{holonomy}
\end{equation}
is called the holonomy associated with the loop. In the case when
the $n$-th energy level is non-degenerate ($d_n=1$), the
corresponding holonomy reduces to the Berry phase \cite{Berry84}.
The holonomy $U_n^{\gamma}$ is a geometric object which is invariant
under gauge transformations corresponding to changing the basis of
the $n$-th eigenspace along the curve $\gamma$. The set
$\textrm{Hol}(A_n)=\{U_n^\gamma| \gamma\in
L_{\lambda_0}(\mathcal{M})\}$, where $L_{\lambda_0}(\mathcal{M})=
\{\gamma: [0,T]\rightarrow \mathcal{M}
|\gamma(0)=\gamma(T)=\lambda_0\}$ is the space of all loops based on
$\lambda_0$, is a subgroup of $U(d_n)$ called the holonomy group.

In Refs.~\cite{ZR99, PZR99} it was shown that if the dimension of
the control manifold is sufficiently large, quantum holonomies can
be used as a means of universal quantum computation. In this
approach, logical states are encoded in the degenerate eigenspace of
a Hamiltonian and gates are implemented by adiabatically varying the
Hamiltonian along suitable loops in the parameter manifold (for a
construction of a universal set of gates, see also
Ref.~\cite{NNS03}).

We point out that many assumptions behind this simple model of HQC
can be relaxed. For example, if we are interested in performing
computation in the $n$-th eigenspace of the Hamiltonian, it is
sufficient that this eigenspace is adiabatically decoupled from the
rest of the Hilbert space, and it is not necessary that there are no
crossings between other energy levels that are separated from the
$n$-th level by energy gaps. Furthermore, in order to obtain a
gauge-invariant expression for the geometric transformation taking
place inside the $n$-th eigenspace, it is not necessary that the
entire Hamiltonian undergoes a cyclic change---it is enough to take
the $n$-th eigenspace around a loop. In fact, the form of the
restriction of the Hamiltonian on the orthogonal complement of that
subspace is irrelevant since what is important for the geometric
transformation taking place inside an adiabatically decoupled
eigenspace is how this subspace changes inside the full Hilbert
space. More precisely, adiabatic quantum holonomies inside the
$n$-th eigenspace can be equivalently understood as arising from
parallel transport of vectors along the tautological bundle whose
base is the Grassmannian parametrizing the set of $d_n$-dimensional
subspaces of the full Hilbert space, rather than from parallel
transport along the corresponding bundle over the full space of
control parameters. We note that even the requirement for a closed
loop in the Grassmannian can be relaxed using the notion of
open-path holonomies \cite{KAS06}. The approach that we pursue in
this paper can be best understood as based on closed loops in the
Grassmannian, even though---with a small modification---it can be
made to be exactly of the type discussed in the above formulation,
where the Hamiltonian family has a fixed degeneracy structure, and
gates are implemented by loops in the control manifold (see
Sec.\ref{Themainidea}).

\subsection{Stabilizer codes and fault-tolerant computation}

A large class of quantum error-correcting codes can be described by
the so called stabilizer formalism \cite{Got96, CRSS96, CRSS96'}. A
stabilizer $S$ is an Abelian subgroup of the Pauli group
$\mathcal{G}_n$ on $n$ qubits that does not contain the element $-I$
\cite{NieChu00}. The Pauli group consists of all possible $n$-fold
tensor products of the Pauli matrices $\sigma^x\equiv X$,
$\sigma^y=Y$, $\sigma^z=Z$ together with the multiplicative factors
$\pm1$, $\pm i$. The stabilizer code corresponding to $S$ is the
subspace of all states $|\psi\rangle$ which are left invariant under
the action of every operator in $S$ ($G|\psi\rangle=|\psi\rangle$,
$\forall G \in S$). It is easy to see that the stabilizer of a code
encoding $k$ qubits into $n$ has $n-k$ generators. Recently, a more
general notion of codes has been introduced---\textit{subsystem} or
\textit{operator} codes \cite{KLP05, KLPL06}---that employs the most
general encoding of information, encoding in subsystems
\cite{Knill:99a, Kempe:01, Knill06}. In the case of a single
subsystem, the Hilbert space decomposes as
$\mathcal{H}=\mathcal{H}^A\otimes \mathcal{H}^B\oplus\mathcal{K}$,
where $\mathcal{H}^A$ is the subsystem in which the logical
information is encoded, $\mathcal{H}^B$ is the \textit{gauge}
subsystem, and $\mathcal{K}$ is the rest of the Hilbert space. For
operator stabilizer codes, the stabilizer leaves the subspace
$\mathcal{H}^A\otimes \mathcal{H}^B$ invariant but the encoded
information is invariant also under operations that act on the gauge
subsystem. An operator stabilizer code encoding $k$ qubits into $n$
with $r$ gauge qubits has $n-r-k$ stabilizer generators, while the
gauge group has $2r$ generators \cite{Pou05}. According to the
error-correction condition for stabilizer codes \cite{NieChu00,
Pou05}, a set of errors $\{E_i\}$ in $\mathcal{G}_n$ (which without
loss of generality are assumed to be Hermitian) is correctable by
the code if and only if, for all $i$ and $j$, $E_iE_j$ anticommutes
with at least one element of $S$, or otherwise belongs to $S$ or to
the gauge group. In this paper we will be concerned with stabilizer
codes for the correction of single-qubit errors and the techniques
for fault-tolerant computation \cite{Shor96, DVS96, ABO98, Kit97,
KLZ98, Got97', Got97, Pre99} on such codes.

A quantum information processing scheme is called fault-tolerant if
a single error occurring during the implementation of any given
operation introduces at most one error per block of the code
\cite{Got97'}. This property has to apply for unitary gates as well
as measurements, including those that constitute the
error-correcting operations themselves. Fault-tolerant schemes for
computation on stabilizer codes generally depend on the code being
used---some codes, such as the Bacon-Shor subsystem codes
\cite{Bac06, BC06}, for example, are better suited for
fault-tolerant computation than others \cite{AC07}. In spite of
these differences, however, it has been shown that fault-tolerant
information processing is possible on any stabilizer code
\cite{Got97', Got97}. The general procedure can be described briefly
as follows. DiVincenzo and Shor \cite{DVS96} demonstrated how to
perform fault-tolerant measurements of the stabilizer for any
stabilizer code. Their method makes use of an approach introduced by
Shor \cite{Shor96}, which involves the ``cat'' state $(|0...0\rangle
+ |1...1\rangle)/\sqrt{2}$ which can be prepared and verified with a
satisfactory precision. As pointed out by Gottesman \cite{Got97'},
by the same method one can measure any operator in the Pauli group.
Since the encoded $X$, $Y$ and $Z$ operators belong to the Pauli
group for any stabilizer code \cite{Got97'}, one can prepare
fault tolerantly various superpositions of the logical basis states
$|\overline{0}\rangle$ and $|\overline{1}\rangle$, such as
$|\overline{+}\rangle=(|\overline{0}\rangle
+|\overline{1}\rangle)/\sqrt{2}$, for example. The latter can be
used to implement fault tolerantly the encoded Phase and Hadamard
gates as long as a fault-tolerant C-NOT gate is available
\cite{Got97'}. Gottesman showed how the C-NOT gate can be
implemented fault tolerantly by first applying a transversal
operation on four encoded qubits and then measuring the encoded $X$
operator on two of them. Finally, for universal computation one
needs a gate outside of the Clifford group, e.g., the Toffoli gate.
The Toffoli construction was demonstrated first by Shor in
Ref.~\cite{Shor96} for a specific type of codes---those obtained
from doubly-even self-dual classical codes by the
Calderbank-Shor-Steane (CSS) construction \cite{CS96, St96b}.
Gottesman showed \cite{Got97} that a transversal implementation of
the same procedure exists for any stabilizer code.

\subsection{Overview of the scheme} \label{Themainidea}

Note that the described method for universal fault-tolerant
computation on stabilizer codes uses almost exclusively transversal
operations---these are operations for which each qubit in a block
interacts only with the corresponding qubit from another block or
from a special ancillary state such as Shor's ``cat" state (see also
Steane's \cite{Ste97} and Knill's \cite{Kni05} methods). However,
transversal operations are not the most general class of operations
that do not lead to propagation of errors. For example, every
transversal operation in a given fault-tolerant protocol can be
substituted by the same operation followed by an operation that
multiplies each syndrome subspace by a different phase, and the
resultant protocol will still be fault-tolerant. This can be easily
seen from the fact that if after a transversal operation the state
is correctable, then it will still be correctable after multiplying
each syndrome subspace by a phase because the correction procedure
involves a projection on one of the syndrome subspaces and thus the
overall phase in that subspace is irrelevant. An operation which is
equal to a transversal operation followed by a transformation on the
gauge subsystem can be similarly seen to be fault-tolerant. It is
these more general fault-tolerant transformations by means of which
we will realize fault-tolerant HQC.

To explain the main idea behind our approach, let us consider the
case of standard (subspace) stabilizer codes first. Our goal will be
to find a holonomic realization of a universal set of encoded gates
by adiabatically transporting the code space along suitable loops
via sequences of elementary fault-tolerant transformations of the
above type. The scheme we will present can be roughly described as
follows. We choose as a starting Hamiltonian an element of the
stabilizer and vary this Hamiltonian in an adiabatic manner along
appropriate paths so that from the point of view of the basis of the
full Hilbert space, the vectors in both eigenspaces of the
Hamiltonian undergo the same transversal transformation. Under this
procedure, each eigenspace will acquire a dynamical phase
corresponding to the energy of that eigenspace, but since the
Hamiltonian is an element of the stabilizer, these phases will
amount to relative phases between different syndrome subspaces and
they would be projected out if a measurement of the syndrome is
performed. So from the point of view of the basis of the full
Hilbert space, the overall transformation under this procedure is of
the general fault-tolerant type we described. Since the code space
is a subspace of an eigenspace of the Hamiltonian, it will be
effectively transformed by the corresponding transversal operation.

The standard fault-tolerant procedures provide prescriptions of how
to implement any encoded gate by a sequence of elementary
transversal operations. Therefore, if we make the code space follow
an appropriate sequence of transversal operations in the described
adiabatic manner, when we complete an encoded operation we will have
taken the code space around a loop whose associated holonomy is
equal to the encoded operation. A simple way to see that this is
indeed a holonomy is to notice that if we track the initial code
space as it evolves, it undergoes a loop in the Grassmannian since
at the end we complete an encoded gate. Furthermore, at all times
the Hamiltonian acts trivially on the code space so that all states
inside it acquire the same dynamical phase. Hence the nontrivial
transformation resulting inside the code space must be geometric.
Thus by following precisely the sequence of transversal operations
and measurements that are used in a given dynamical fault-tolerant
scheme, we obtain a scheme that implements logical gates through
adiabatic holonomies and at the same time is fault-tolerant.

In the case of subsystem codes, the code subsystem can be thought of
as a collection of subspaces, each of which contains the same
redundant information. The relative phases between each of these
subspaces are gauge degrees of freedom. Applying a particular
encoded gate is equivalent to applying the same gate in each
subspace. In this case, our scheme can use Hamiltonians that are
either elements of the stabilizer or elements of the gauge group. If
the Hamiltonian is an element of the stabilizer, all ``redundant''
subspaces inside a given syndrome subspace belong to a single
eigenspace of the Hamiltonian, and all of them will undergo the
transversal operation that effectively transforms that eigenspace.
If the Hamiltonian is an element of the gauge group, then some of
the subspaces of interest will belong to the ground space while
others will belong to the excited space. However, since the scheme
implements the same transversal operation in each eigenspace, all
subspaces by construction will undergo the same transversal
operation. The relative dynamical phases acquired between subspaces
in the ground and excited eigenspaces amount to a gauge
transformation that does not affect the fault tolerance of the
operation. Thus by following the same sequence of transversal
operations and measurements as in a standard dynamical scheme, we
obtain a fault-tolerant holonomic realization of encoded computation
on subsystem codes.

From the point of view of the full Hilbert space, this method
performs transformations of the generalized fault-tolerant type we
described earlier.  This is what ensures the fault tolerance of the
method. From a geometrical point of view, this corresponds to
transporting each of the different syndrome subspaces around a loop
such that all of them simultaneously undergo the same geometric
transformation (in the case of subsystem codes, the same applies for
all redundant subspaces inside each syndrome space). The statement
that all subspaces simultaneously undergo the same geometric
transformation makes sense with respect to a particular choice of
the instantaneous basis in each subspace. For any choice of basis in
the code space at a given point along the loop, there is a preferred
choice of bases inside the rest of the syndrome spaces that is
determined by the notion of correctable errors. Correctable errors
can be thought of as transitions from the code space to the other
syndrome spaces which can be undone if error correction is applied.
These errors therefore map the basis from the code space to
particular bases in the error spaces such that with respect to these
bases, a state that has undergone a correctable error has the same
form as the form that the non-erroneous state has with respect to
the basis of the code space. Our scheme applies the same geometric
transformation, in this sense, in all syndrome subspaces.

Note that the present approach differs from the original HQC model
\cite{ZR99, PZR99} in that it computes in several subspaces at the
same time. Another difference from the original model is that we do
not use a single family of iso-degenerate Hamiltonians. This is
because for simplicity we use Hamiltonians that are equal to a
single element of the stabilizer or the gauge group at a given time,
and we change the Hamiltonians along different portions of the loop.
Thus, if the Hamiltonians are elements of the stabilizer, a
particular syndrome subspace may belong to the ground space of the
Hamiltonian during one portion of the loop and to the excited space
during another one. Therefore, the holonomies in our scheme can be
most naturally understood as resulting from parallel transport along
loops in the tautological bundle over the Grassmannian (this is the
fiber bundle whose base is the Grassmannian and whose fibers are the
subspaces corresponding to the different points in the
Grassmannian), rather than loops in a bundle over a space
parametrizing an iso-degenerate family of Hamiltonians. In the case
of subsystem codes, if along the loop we change between Hamiltonians
which are non-commuting elements of the gauge group, the redundant
subspaces that constitute a logical subsystem may seem to undergo
dynamical transformations in addition to the geometric ones.
However, these dynamical changes are equivalent to gauge
transformations and do not affect the workings of the scheme. We
could modify our scheme so that it uses a Hamiltonian that separates
all of the subspaces of interest by energetic gaps and adiabatically
transports each of these subspaces along the same path that it would
follow under the scheme we described. Then the holonomy resulting in
each eigenspace could be understood as being of the original HQC
type. However, this is unnecessary since the dynamical phases are
irrelevant for the workings of our model.

As single-qubit unitaries together with the C-NOT gate form a
universal set of gates, fault-tolerant computation can be realized
entirely in terms of single-qubit operations and C-NOT operations
between qubits from different blocks, assuming that the ``cat''
state can be prepared reliably. Hence our task will be to find
adiabatic realizations of these operations, as well as of the
operations for preparing and verifying the ``cat'' state. Then these
operations can be used as building blocks to implement
fault-tolerant HQC according to the idea described in this section.

\section{The scheme}

Consider a $[[n,1,r,3]]$ stabilizer code. This is a code that
encodes $1$ qubit into $n$, has $r$ gauge qubits, and can correct
arbitrary single-qubit errors. To perform a holonomic operation on
this code, we need a nontrivial starting Hamiltonian that leaves the
code space or code subsystem invariant. It is easy to verify that
the only Hamiltonians that satisfy this property are linear
combinations of the elements of the stabilizer and, in the case of
subsystem codes, elements of the gauge group.

Note that the stabilizer and the gauge group transform during the
course of the computation under the operations being applied. At any
stage when we complete an encoded operation, they return to their
initial forms. Our scheme will follow the same transversal
operations as those used in a standard dynamical fault-tolerant
scheme, but as we explained in the previous section, in addition we
will have extra dynamical phases that multiply each syndrome
subspace or are equivalent to gauge transformations. However, it is
easy to see that these phases do not affect the way the stabilizer
or the gauge group transform, so we can omit them from our analysis
of the transformation of these groups.

During the implementation of a standard encoded gate, the Pauli
group $\mathcal{G}_n$ on a given codeword may change in such a way that it acts on other codewords, but it can be verified that this ``spreading'' can be limited
to at most $4$ other codewords including the ``cat" state. This is
because the encoded C-NOT gate can be implemented fault tolerantly
on any stabilizer code by a transversal operation on $4$ encoded
qubits \cite{Got97}, and any encoded Clifford gate can be realized
using only the encoded C-NOT, provided that we are able to do
fault-tolerant measurements (the encoded Clifford group is generated
by the encoded Hadamard, Phase and C-NOT gates). Encoded gates
outside of the Clifford group, such as the encoded $\pi/8$ or
Toffoli gates, can be implemented fault tolerantly using encoded
C-NOT gates conditioned on the qubits in a ``cat" state, so they may
require transversal operations on a total of $5$ blocks. More
precisely, the fault-tolerant implementation of the Toffoli gate
requires the preparation of a special state of three encoded qubits
\cite{Shor96}, which involves a sequence of conditional encoded
Phase operations and conditional encoded C-NOT operations with
conditioning on the qubits in a ``cat" state \cite{Got97}. But the
encoded Phase gate has a universal implementation using an encoded
C-NOT between the qubit and an ancilla, so the conditional Phase
gate may require applying a conditional encoded C-NOT. The procedure
for implementing an encoded $\pi/8$ gate involves applying an
encoded $SX$ gate conditioned on the qubits in a ``cat" state
\cite{BMPRV99} ($S$ denotes the Phase gate), but the encoded $S$
gate generally also involves an encoded C-NOT on the qubit and an
ancilla, so it may also require the interaction of $4$ blocks. For
CSS codes, however, the spreading of the Pauli group that acts on a given block can be
limited to a total of $3$ blocks during the implementation of a basic encoded operation since the encoded C-NOT gate has a
transversal implementation \cite{Got97}.

It also should be noted that fault-tolerant encoded Clifford
operations can be implemented using only Clifford gates on the
physical qubits \cite{Got97}. These operations transform the
stabilizer and the gauge group into subgroups of the Pauli group,
and their elements remain in the form of tensor products of Pauli
matrices. The fault-tolerant implementation of encoded gates outside
of the Clifford group, however, involves operations that take these
groups outside of the Pauli group. We will, therefore, consider
separately two cases: encoded operations in the Clifford group, and
encoded operations outside of the Clifford group.

\subsection{Encoded operations in the Clifford group}

In Ref.~\cite{Got97} it was shown that every encoded operation in
the Clifford group can be implemented fault tolerantly using
Clifford gates on physical qubits. The Clifford group is generated
by the Hadamard, Phase and C-NOT gates, but in addition to these
gates, we will also demonstrate the adiabatic implementation of the
$X$ and $Z$ gates which are standard for quantum computation. We
will restrict our attention to implementing single-qubit unitaries
on the first qubit in a block, as well as C-NOT operations between
the first qubits in two blocks. The operations on the rest of the
qubits can be obtained analogously.

\subsubsection{Single-qubit unitary operations}

In order to implement a single-qubit operation on the first qubit in
a block, we will choose as a starting Hamiltonian an element of the
stabilizer (with a minus sign) or an element of the gauge group that
acts non-trivially on that qubit. Since we are considering codes
that can correct arbitrary single-qubit errors, one can always find
an element $\widehat{G}$ of the initial stabilizer or the initial gauge group that
has a factor $\sigma^0=I$, $\sigma^1=X$, $\sigma^2=Y$ or
$\sigma^3=Z$ acting on the first qubit, i.e.,
\begin{equation}
\widehat{G}=\sigma^i\otimes \widetilde{G},\hspace{0.4cm}i=0,1,2,3,
\label{stabelementgen}
\end{equation}
where $\widetilde{G}$ is a tensor product of Pauli matrices and the
identity on the remaining $n-1$ qubits. It can be verified that
under Clifford gates the stabilizer and the gauge group transform in
such a way that this is always the case except that the factor
$\widetilde{G}$ may spread to qubits in other blocks. From now on,
we will use ``hat" to denote operators on all these qubits and
``tilde" to denote operators on all the qubits except the first one.

Without loss of generality we will assume that the chosen stabilizer
or gauge-group element for that qubit has the form
\begin{equation}
\widehat{G}=Z\otimes \widetilde{G}.\label{stabelement}
\end{equation}
As initial Hamiltonian, we will take the operator
\begin{equation}
\widehat{H}(0)=-\widehat{G}=-Z\otimes \widetilde{G}.\label{H(0)}
\end{equation}
Thus, if $\widehat{G}$ is an element of the stabilizer, the code
space will belong to the ground space of $\widehat{H}(0)$. Our goal
is to find paths in the space of parameters of the Hamiltonian such
that when the Hamiltonian is varied adiabatically along these paths,
each of its eigenspaces undergoes a transformation which is
equivalent to that caused by a single-qubit operation on the first
qubit.

\textit{Proposition 1.} If the initial Hamiltonian \eqref{H(0)} is
varied adiabatically so that only the factor acting on the first
qubit changes,
\begin{equation}
\widehat{H}(t)=-H(t)\otimes \widetilde{G},\label{Ham1}
\end{equation}
where
\begin{equation}
\textrm{Tr}\{H(t)\}=0,
\end{equation}
the transformation that each of the eigenspaces of this Hamiltonian
undergoes will be equivalent to that resulting from a local unitary
on the first qubit up to a global phase, i.e., the geometric part
$\widehat{U}_g(t)=U^{\lambda}_{A_0}\oplus U^{\lambda}_{A_1}$ of the
overall unitary, where $U^{\lambda}_{A_i}$, $i=0,1$, are the
path-ordered exponents \eqref{openpathholonomy} corresponding to the
ground and excited spaces, respectively, will be equal to a local
unitary on the first qubit, $\widehat{U}_g(t)= U(t)\otimes \widetilde{I}$.\\

\textit{Proof.} Observe that \eqref{Ham1} can be written as
\begin{equation}
\widehat{H}(t)=H(t)\otimes \widetilde{P}_0- H(t)\otimes
\widetilde{P}_1,\label{Ham2}
\end{equation}
where
\begin{equation}
\widetilde{P}_{0}=\frac{\widetilde{I}- \widetilde{G}}{2},
\hspace{0.3cm}\widetilde{P}_{1}=\frac{\widetilde{I}+
\widetilde{G}}{2}, \label{Projectors}
\end{equation}
are orthogonal complementary projectors. The evolution driven by
$\widehat{H}(t)$ is therefore
\begin{equation}
\widehat{U}(t) = U_0(t)\otimes \widetilde{P}_0 + U_1(t) \otimes
\widetilde{P}_1,\label{overallunitary}
\end{equation}
where
\begin{equation}
U_{0}(t)=\mathcal{T}\textrm{exp}(-i\overset{t}{\underset{0}{\int}}
H(\tau)d\tau), \hspace{0.3cm}
U_{1}(t)=\mathcal{T}\textrm{exp}(i\overset{t}{\underset{0}{\int}}
H(\tau)d\tau).\label{unitaries}
\end{equation}

Let $|\phi_{0}(t)\rangle$ and $|\phi_{1}(t)\rangle$ be the
instantaneous ground and excited states of $H(t)$ with eigenvalues
$E_{0}(t)=- E(t)$, $E_{1}(t)= E(t)$ ($E(t)>0$). Using
Eq.~\eqref{adiabaticevolution} for the expressions
\eqref{unitaries}, we obtain that in the adiabatic limit,
\begin{equation}
U_{j}(t)= e^{i\omega(t)}U_{A_{j}}(t)\oplus
e^{-i\omega(t)}U_{A_{j}}(t),\hspace{0.3cm}j=0,1,\label{unita}
\end{equation}
where $\omega(t)= \int_0^t d\tau E(\tau)$ and
\begin{gather}
U_{A_{j}}(t)=e^{\int_0^t d\tau
\langle\phi_{j}(\tau)|\frac{d}{d\tau}|\phi_{j}(\tau)\rangle
}|\phi_{j}(t)\rangle \langle \phi_{j}(0)|,
\hspace{0.3cm}j=0,1.\label{holon}
\end{gather}

The projectors on the ground and excited eigenspaces of
$\widehat{H}(0)$ are
\begin{equation}
\widehat{P}_0=|\phi_0(0)\rangle\langle\phi_0(0)|\otimes
\widetilde{P}_0 + |\phi_1(0)\rangle\langle\phi_1(0)|\otimes
\widetilde{P}_1
\end{equation}
and
\begin{equation}
\widehat{P}_1=|\phi_1(0)\rangle\langle\phi_1(0)|\otimes
\widetilde{P}_0 + |\phi_0(0)\rangle\langle\phi_0(0)|\otimes
\widetilde{P}_1,
\end{equation}
respectively. Using Eq.~\eqref{unita} and Eq.~\eqref{holon}, one
can see that the effect of the unitary \eqref{overallunitary} on
each of these projectors is
\begin{equation}
\widehat{U}(t)\widehat{P}_0=e^{i\omega(t)}(U_{A_{0}}(t)\oplus
U_{A_{1}}(t))\otimes\widetilde{I} \hspace{0.1cm}\widehat{P}_0,
\end{equation}
\begin{equation}
\widehat{U}(t)\widehat{P}_1=e^{-i\omega(t)}(U_{A_{0}}(t)\oplus
U_{A_{1}}(t))\otimes\widetilde{I} \hspace{0.1cm}\widehat{P}_1,
\end{equation}
i.e, up to an overall dynamical phase its effect on each of the
eigenspaces is the same as that of the unitary
\begin{equation}
\widehat{U}_g(t)=U(t)\otimes\widetilde{I},
\end{equation}
where
\begin{equation}
U(t)=U_{A_{0}}(t)\oplus U_{A_{1}}(t)\label{finalU}.
\end{equation}
This completes the proof.

We next show how by suitably choosing $H(t)$ we can implement all
necessary single-qubit gates. We will identify a set of points in
parameter space, such that by interpolating between these points we
can draw various paths resulting in the desired transformations. We
remark that if a path does not form a loop, the geometric
transformation \eqref{finalU} could be associated with an open-path
holonomy taking place inside each eigenspace of the Hamiltonian,
provided that the final eigenspaces have non-zero overlap with the
corresponding initial ones \cite{KAS06} (see
Appendix~\ref{AppendixB}).

Consider the single-qubit unitary operator
\begin{equation}
V^{\theta\pm}=\frac{1}{\sqrt{2}}\begin{pmatrix} 1& \mp e^{-i\theta}\\
\pm e^{i\theta}&1
\end{pmatrix},
\end{equation}
where $\theta$ is a real parameter. Note that $V^{\theta
\mp}=(V^{\theta \pm})^{\dagger}$. Define the following single-qubit
Hamiltonian:
\begin{equation}
H^{\theta\pm}\equiv V^{\theta\pm}ZV^{\theta\mp}.
\end{equation}
Let $H(t)$ in Eq.~\eqref{Ham1} be a Hamiltonian which interpolates
between $H(0)=Z$ and $H(T)=H^{\theta\pm}$ (up to a factor) as
follows:
\begin{equation}
H(t)=f(t)Z+g(t) H^{\theta\pm}\equiv H^{\theta\pm}_{f,g}(t),
\label{interpolation}
\end{equation}
where $f(0)>0$, $g(T)>0$, and $f(T)=g(0)=0$. To simplify our
notation, we will drop the indices $f$ and $g$ of the Hamiltonian,
since the exact form of these functions is not important for our
analysis as long as they are sufficiently smooth (see discussion
below). This Hamiltonian has eigenvalues $\pm \sqrt{f(t)^2+g(t)^2}$
and its energy gap is non-zero unless the entire Hamiltonian
vanishes.

\textit{Proposition 2.} In the adiabatic limit, the Hamiltonian
\eqref{Ham1} with $H(t)=H^{\theta\pm}(t)$ gives rise to the
geometric transformation $V^{\theta\pm}\otimes \widetilde{I}$.

The proof of this proposition is presented in Appendix A.

We will use this result to construct a set of standard gates by
sequences of operations of the form $V^{\theta\pm}$, which can be
generated by interpolations of the type \eqref{interpolation} run
forward or backward. For single-qubit gates in the Clifford group,
we will only need three values of $\theta$: $0$, $\pi/2$ and
$\pi/4$. For completeness, however, we will also demonstrate how to
implement the $\pi/8$ gate, which together with the Hadamard gate is
sufficient to generate any single-qubit unitary transformation
\cite{BMPRV99}. For this we will need $\theta=\pi/8$. Note that
\begin{equation}
H^{\theta\pm}= \pm (\cos{\theta}X+\sin{\theta}Y),
\end{equation}
so for these values of $\theta$ we have $H^{0\pm}=\pm X$,
$H^{\pi/2\pm}=\pm Y$, $H^{\pi/4\pm}=\pm
(\frac{1}{\sqrt{2}}X+\frac{1}{\sqrt{2}}Y)$, $H^{\pi/8\pm}=\pm
(\cos{\frac{\pi}{8}}X+\sin{\frac{\pi}{8}}Y)$.

Consider the following adiabatic interpolations:
\begin{equation}
-Z\otimes\widetilde{G} \rightarrow -Y\otimes \widetilde{G}
\rightarrow Z\otimes\widetilde{G}.\label{Xgate}
\end{equation}
According to the above result, the first interpolation yields the
transformation $V^{\pi/2+}$. The second interpolation can be
regarded as the inverse of $Z\otimes\widetilde{G}\rightarrow
-Y\otimes\widetilde{G}$ which is equivalent to
$-Z\otimes\widetilde{G}\rightarrow Y\otimes\widetilde{G}$ since
$\widehat{H}(t)$ and $-\widehat{H}(t)$ yield the same geometric
transformations. Thus the second interpolation results in
$(V^{\pi/2-})^{\dagger}=V^{\pi/2+}$. The net result is therefore
$V^{\pi/2+}V^{\pi/2+}=iX$. We see that up to a global phase, the
geometric part of the transformation resulting from the above
sequence is equal to the single-qubit $X$ gate.

Similarly, one can verify that the $Z$ gate can be realized via
the loop
\begin{equation}
-Z\otimes\widetilde{G} \rightarrow -X\otimes
\widetilde{G}\rightarrow Z\otimes\widetilde{G} \rightarrow
Y\otimes \widetilde{G}\rightarrow
-Z\otimes\widetilde{G}.\label{Zgate}
\end{equation}

The Phase gate can be realized by applying
\begin{equation}
-Z\otimes\widetilde{G}\rightarrow
-(\frac{1}{\sqrt{2}}X+\frac{1}{\sqrt{2}}Y)\otimes\widetilde{G}\rightarrow
Z\otimes\widetilde{G},
\end{equation}
followed by the $X$ gate.

The Hadamard gate can be realized by first
applying $Z$, followed by
\begin{equation}
-Z\otimes\widetilde{G} \rightarrow -X\otimes \widetilde{G}.
\end{equation}

Finally, the $\pi/8$ gate can be implemented by first applying
$Y=iXZ$, followed by
\begin{equation}
Z\otimes\widetilde{G}\rightarrow
-(\cos{\frac{\pi}{8}}X+\sin{\frac{\pi}{8}}Y)\otimes\widetilde{G}\rightarrow
-Z\otimes\widetilde{G}.
\end{equation}

\textit{Comment.} We emphasize that the elementary gates we
construct here are not holonomies associated with the code space or
the error subspaces. The holonomy in a given subspace is a
transformation whose components are defined with respect to a basis
of that subspace, while these elementary gates are the geometric
parts of the unitaries resulting in the full Hilbert space under the
described adiabatic evolutions. As explained in Sec. II C, only
after we compose a suitable sequence of such elementary gates do we
obtain a closed-loop holonomy in the code space (or the code
subsystem). The non-Abelian holonomies through which we perform HQC
in the code space and the error subspaces are the universal set of
encoded gates that we would obtain by composing the set of
elementary one- and two-qubit gates (combined with measurements if
necessary) according to the rules of a given dynamical
fault-tolerant scheme.

\subsubsection{Note on the adiabatic condition}

Before we show how to implement the C-NOT gate between two qubits,
let us comment on the conditions under which the adiabatic
approximation assumed in the above operations is satisfied. Because
of the form \eqref{overallunitary} of the overall unitary, the
adiabatic approximation depends on the extent to which each of the
unitaries \eqref{unitaries} approximates the expression
\eqref{unita}. The latter depends only on the adiabatic properties
of the non-degenerate two-level Hamiltonian $H(t)$. For such a
Hamiltonian, the simple version of the adiabatic condition
\cite{mes} reads
\begin{equation}
\frac{\varepsilon}{{\Delta}^2}\ll 1, \label{adi}
\end{equation}
where
\begin{equation} \varepsilon =
\operatornamewithlimits{max}_{0\leq t\leq T}|\langle
\phi_1(t)|\frac{dH(t)}{dt}|\phi_0(t)\rangle|, \label{eps}
\end{equation}
and
\begin{equation}
\Delta = \operatornamewithlimits{min}_{0\leq t\leq
T}(E_1(t)-E_0(t))= \operatornamewithlimits{min}_{0\leq t\leq
T}2E(t) \label{Delt}
\end{equation}
is the minimum energy gap of $H(t)$.

Along the segments of the parameter paths we described, the
Hamiltonian is of the form \eqref{interpolation} and its derivative
is
\begin{equation}
\frac{dH^{\theta\pm}(t)}{dt}=\frac{df(t)}{dt}Z+\frac{dg(t)}{dt}H^{\theta\pm},
\hspace{0.6cm} 0<t<T.
\end{equation}
This derivative is well defined as long as $\frac{df(t)}{dt}$ and
$\frac{dg(t)}{dt}$ are well defined. The curves we described,
however, may not be differentiable at the points connecting two
segments. In order for the Hamiltonians \eqref{interpolation} that
interpolate between these points to be differentiable, the functions
$f(t)$ and $g(t)$ have to satisfy $\frac{df(T)}{dt}=0$ and
$\frac{dg(0)}{dt}=0$. This means that the change of the Hamiltonian
slows down to zero at the end of each segment (except for a possible
change in its strength), and increases again from zero along the
next segment. We point out that when the Hamiltonian stops changing,
we can turn it off completely by decreasing its strength. This can
be done arbitrarily fast and it would not affect a state which
belongs to an eigenspace of the Hamiltonian. Similarly, we can turn
on another Hamiltonian for the implementation of a different
operation.

The above condition guarantees that the adiabatic approximation is
satisfied with precision
$\textit{O}((\frac{\varepsilon}{{\Delta}^2})^2)$. It is known,
however, that under certain conditions on the Hamiltonian, we can
obtain better results \cite{HJ02, HL08}. Let us write the
Schr\"{o}dinger equation as
\begin{equation}
i\frac{d}{dt}|\psi(t)\rangle = H(t) |\psi(t)\rangle\equiv
\frac{1}{\epsilon} \bar{H}(t) |\psi(t)\rangle,
\end{equation}
where $\epsilon>0$ is small. Assume that $\bar{H}(t)$ is smooth and
all its derivatives vanish at the end points $t=0$ and $t=T$ (note
that this means that $\bar{H}(t)$ is non-analytic at these points,
unless it is constant; an alternative strategy is to consider
analytic Hamiltonians with a finite number of vanishing derivatives
at $t=0,T$ \cite{HL08}). Then if we keep $\bar{H}(t)$ fixed and vary
$\epsilon$, the adiabatic error would scale super-polynomially with
$\epsilon$, i.e., the error will decrease with $\epsilon$ faster
than $\textit{O}(\epsilon^N)$ for any $N$ \cite{HJ02}. (Notice that
$\frac{\varepsilon}{{\Delta}^2}\propto \epsilon$, i.e., the error
according to the standard adiabatic approximation is
$\textit{O}(\epsilon^2)$.)

In our case, the smoothness condition translates directly to the
functions $f(t)$ and $g(t)$. For any smooth $f(t)$ and $g(t)$ we can
further ensure that the condition at the end points is satisfied by
the reparametrization $f(t)\rightarrow f(y(t))$, $g(t)\rightarrow
g(y(t))$ where $y(t)$ is a smooth function of $t$ which satisfies
$y(0)=0$, $y(T)=T$, and has vanishing derivatives at $t=0$ and
$t=T$. Then by slowing down the change of the Hamiltonian by a
constant factor $\epsilon$, which amounts to increasing the
total time $T$ by a factor $1/\epsilon$, we can decrease the error
super-polynomially in $\epsilon$. We will use this result to obtain
a low-error interpolation in Sec. IV, where we estimate the time
needed to implement a geometric gate with a certain precision.

\subsubsection{The C-NOT gate}

The stabilizer (or gauge group) on multiple blocks of the code is a
direct product of the stabilizers (or gauge groups) of the
individual blocks. Therefore, from Eq.~\eqref{stabelementgen} it
follows that one can always find an element of the initial
stabilizer or gauge group on multiple blocks that has any desired
combination of factors $\sigma^i$, $i=0,1,2,3$, on the first qubits
in these blocks. It can be verified that applying transversal
Clifford operations on the blocks does not change this property.
Therefore, we can find an element of the stabilizer or the gauge
group that has the form \eqref{stabelement}, where the factor $Z$
acts on the target qubit and $\widetilde{G}$ acts trivially on the
control qubit. We now explain how to implement the C-NOT gate
geometrically starting from such a Hamiltonian.

Notice that a Hamiltonian of the form
\begin{equation}
\widehat{\widehat{H}}(t)=|0\rangle\langle 0|^c \otimes
{H}_{0}(t)\otimes\widetilde{G}+|1\rangle\langle 1|^c\otimes
{H}_{1}(t)\otimes\widetilde{G},\label{hathatH}
\end{equation}
where the superscript $c$ denotes the control qubit, gives rise to
the unitary transformation
\begin{equation}
\widehat{\widehat{U}}(t)=|0\rangle\langle 0|^c \otimes
\widehat{U}_0(t)+|1\rangle\langle 1|^c\otimes
\widehat{U}_1(t),\label{doublehatUU}
\end{equation}
where
\begin{equation}
\widehat{U}_{0,1}(t)=\mathcal{T}\textrm{exp}(-i\overset{t}{\underset{0}{\int}}d\tau
H_{0,1}(\tau)\otimes\widetilde{G}).
\end{equation}
If $H_0(t)$ and $H_1(t)$ have the same non-degenerate instantaneous
spectra, and $\textrm{Tr}\{H_{0,1}(t)\}=0$, then from
Eq.~\eqref{adiabaticevolution} and Proposition 1 it follows that in
the adiabatic limit each of the eigenspaces of
$\widehat{\widehat{H}}(t)$ will undergo the geometric transformation
\begin{equation}
\widehat{\widehat{U}}_g(t)=|0\rangle\langle 0|^c \otimes
V_0(t)\otimes \widetilde{I}+|1\rangle\langle 1|^c\otimes
V_1(t)\otimes \widetilde{I},\label{CNOTtransformation}
\end{equation}
where $V_{0,1}(t)\otimes \widetilde{I}$ are the geometric
transformations generated by ${H}_{0,1}(t)\otimes\widetilde{G}$
according to Proposition 1.

Our goal is to find $H_0(t)$ and $H_1(t)$, $H_0(0)=H_1(0)=Z$, such
that at the end of the transformation, the geometric unitary
\eqref{CNOTtransformation} will be equal to the C-NOT gate. In other
words, we want $V_0(2T)=I$ and $V_1(2T)=X$ (here we have chosen the
total time of evolution to be $2T$ for convenience).

We already saw how to generate geometrically the $X$ gate up to a
phase---Eq.~\eqref{Xgate}. We can use the same Hamiltonian in place
of $H_1(t)$:
\begin{equation}
H_1(t)=\begin{cases} H^{\pi/2+}(t), \hspace{0.2cm} 0\leq t \leq T\\
H^{\pi/2-}(2T-t), \hspace{0.2cm} T\leq t \leq 2T.
\end{cases}\label{H1t}
\end{equation}

Now we want to find a Hamiltonian $H_0(t)$ with the same spectrum as
$H_1(t)$, which gives rise to a trivial geometric transformation,
$V_0(2T)=I$ (possibly up to a phase, which can be undone later).
Since all Hamiltonians of the type $H^{\theta\pm}(t)$ have the same
instantaneous spectrum (for fixed $f(t)$ and $g(t)$), we can simply
choose
\begin{equation}
H_0(t)=\begin{cases} H^{\pi/2+}(t), \hspace{0.2cm} 0\leq t \leq T\\
H^{\pi/2+}(2T-t), \hspace{0.2cm} T\leq t \leq 2T,
\end{cases}\label{H0t}
\end{equation}
which corresponds to applying a given transformation from $t=0$ to
$t=T$ and then undoing it (running it backwards) from $t=T$ to
$t=2T$. This results exactly in $V_0(2T)=I$.

Since, as we saw in Sec.~III.A.1, the Hamiltonian $H_1(t)\otimes
\widetilde{G}$ gives rise to the geometric transformation
$iX\otimes\widetilde I$, the above choice for the Hamiltonians
\eqref{H0t} and \eqref{H1t} in Eq.~\eqref{hathatH} will result in
the geometric transformation
\begin{equation}
|0\rangle\langle 0|^c \otimes
I\otimes\widetilde{I}+i|1\rangle\langle 1|^c\otimes
{X}\otimes\widetilde{I},
\end{equation}
which is the desired C-NOT gate up to a Phase gate on the control
qubit. We can correct the phase by applying the inverse of the Phase
gate to the control qubit, either before or after the described
transformation.

Notice that from $t=0$ to $t=T$ the Hamiltonians \eqref{H1t} and
\eqref{H0t} are identical, i.e., during this period the Hamiltonian
\eqref{hathatH} has the form
\begin{equation}
I^c\otimes H^{\pi/2+}(t)\otimes\widetilde{G},
\end{equation}
so we are simply applying the single-qubit operation $V^{\pi/2+}$ to
the target qubit according to the method for single-qubit gates
described before. It is straightforward to verify that during the
second period, from $t=T$ to $t=2T$, the Hamiltonian \eqref{hathatH}
realizes the interpolation
\begin{equation}
-I^c\otimes Y\otimes \widetilde{G} \rightarrow -Z^c\otimes Z\otimes
\widetilde{G},\label{CNOTint}
\end{equation}
which is understood as in Eq.~\eqref{interpolation}.

To summarize, the C-NOT gate can be implemented by first applying
the inverse of the Phase gate ($S^{\dagger}$) on the control qubit,
as well as the transformation $V^{\pi/2+}$ on the target qubit,
followed by the transformation \eqref{CNOTint}. Due to the form
\eqref{hathatH} of $\widehat{\widehat{H}}(t)$, the extent to which
the adiabatic approximation is satisfied during this transformation
depends only on the adiabatic properties of the single-qubit
Hamiltonians $H^{\pi/2\pm}(t)$ which we discussed in the previous
section.

Our construction allowed us to prove the resulting geometric
transformations without explicitly calculating the path-ordered
integrals \eqref{openpathholonomy}. It may be instructive, however,
to demonstrate this calculation for at least one of the gates we
described. In Appendix B, we present an explicit calculation of the
geometric transformation for the $Z$ gate for the following two
cases: $f(t)=1-\frac{t}{T}$, $g(t)=\frac{t}{T}$ (linear
interpolation); $f(t)=\cos{\frac{\pi t}{2 T}}$, $g(t)=\sin
{\frac{\pi t}{2 T}}$ (unitary interpolation).

\subsection{Encoded operations outside of the Clifford group}

For universal fault-tolerant computation we also need at least one
encoded gate outside of the Clifford group. The fault-tolerant
implementation of such gates is based on the preparation of a
special encoded state \cite{Shor96,KLZ98,Got97, BMPRV99,ZLC00} which
involves a measurement of an encoded operator in the Clifford group.
For example, the $\pi/8$ gate requires the preparation of the
encoded state
$\frac{|0\rangle+\textrm{exp}(i\pi/4)|1\rangle}{\sqrt{2}}$, which
can be realized by measuring the encoded operator $e^{-i\pi/4}SX$
\cite{BMPRV99}. Equivalently, the state can be obtained by applying
the encoded operation $RS^{\dagger}$, where $R$ denotes the Hadamard
gate, on the encoded state
$\frac{\cos(\pi/8)|0\rangle+\sin(\pi/8)|1\rangle}{\sqrt{2}}$ which
can be prepared by measuring the encoded Hadamard gate \cite{KLZ98}.
The Toffoli gate requires the preparation of the three-qubit encoded
state $\frac{|000\rangle+|010\rangle+|100\rangle+|111\rangle}{2}$
and involves a similar procedure \cite{ZLC00}. In all these
instances, the measurement of the encoded Clifford operator is
realized by applying transversally the operator conditioned on the
qubits in a ``cat" state.

We now show a general method that can be used to implement
geometrically any conditional transversal Clifford operation with
conditioning on the ``cat" state. Let $O$ be a Clifford gate acting
on the first qubits from some set of blocks. As we discussed in the
previous section, under this unitary the stabilizer and the gauge
group transform in such a way that we can always find an element
with an arbitrary combination of Pauli matrices on the first qubits.
If we write this element in the form
\begin{equation}
\widehat{G}=G_1\otimes G_{2,...,n},
\end{equation}
where $G_1$ is a tensor product of Pauli matrices acting on the
first qubits from the blocks, and $G_{2,...,n}$ is an operator on
the rest of the qubits, then applying $O$ conditioned on the first
qubit in a ``cat" state transforms this stabilizer or gauge-group
element as follows:
\begin{equation}
I^c\otimes G_1\otimes G_{2,...,n}=|0\rangle\langle 0|^c\otimes
G_1\otimes G_{2,...,n}+|1\rangle\langle 1|^c\otimes G_1\otimes
G_{2,...,n} \rightarrow |0\rangle\langle 0|^c\otimes G_1\otimes
G_{2,...,n}+|1\rangle\langle 1|^c\otimes OG_1O^{\dagger}\otimes
G_{2,...,n},
\end{equation}
where the superscript $c$ denotes the control qubit from the ``cat"
state. We can implement this operation by choosing the factor $G_1$
to be the same as the one we would use if we wanted to implement the
operation $O$ according to the previously described procedure. Then
we can apply the following Hamiltonian:
\begin{equation}
\widehat{\widehat{H}}_{C(O)}(t)=-|0\rangle\langle 0|^c\otimes
G_1\otimes G_{2,...,n}-\alpha(t)|1\rangle\langle 1|^c\otimes
H_O(t)\otimes G_{2,...,n},\label{HamA}
\end{equation}
where $-H_O(t)\otimes G_{2,...,n}$ is the Hamiltonian that we would
use for the implementation of the operation $O$, and $\alpha(t)$ is
a real parameter chosen such that at every moment the operator
$\alpha(t)|1\rangle\langle 1|^c\otimes H_O(t)\otimes G_{2,...,n}$
has the same instantaneous spectrum as the operator
$|0\rangle\langle 0|^c\otimes G_1\otimes G_{2,...,n}$. This
guarantees that the overall Hamiltonian is degenerate and the
geometric transformation of each of its eigenspaces is given by the
operator
\begin{equation}
\widehat{\widehat{U}}_g(t)=|0\rangle\langle 0|^c\otimes I_1\otimes
I_{2,...,n}+|1\rangle\langle 1|^c\otimes U_O(t)\otimes
I_{2,...,n},
\end{equation}
where $U_O(t)$ is the geometric transformation on the first qubits
generated by $-H_O(t)\otimes G_{2,...,n}$. Since we presented the
constructions of our basic Clifford operations up to an overall
phase, the operation $U_O(t)$ may differ from the desired operation
by a phase. This phase can be corrected by applying a suitable gate
on the control qubit from the ``cat" state (we explain how this can
be done in the next section). We remark that a Hamiltonian of the
type \eqref{HamA} requires fine tuning of the parameter $\alpha(t)$
and generally can be complicated. Our goal in this section is to
prove that universal fault-tolerant holonomic computation is
possible in principle. In Sec. V we show that depending on the code
one can find more natural implementations of these operations.

If we want to apply a second conditional Clifford operation $Q$ on
the first qubits in the blocks, we can do this as follows. Imagine
that if we had to apply the operation $Q$ following the operation
$O$, we would use the Hamiltonian $\widehat{H}_Q(t)=-H_Q(t)\otimes
G'_{2,...,n}$, where $\widehat{H}_Q(0)=OG'_1O^{\dagger}\otimes
G'_{2,...,n}$ is a suitable element of the stabilizer or the gauge
group after the application of $O$. Before the application of $O$,
that element would have had the form $G'_1\otimes G'_{2,...,n}$.
Under the application of a conditional $O$, the element $G'_1\otimes
G'_{2,...,n}$ transforms to $|0\rangle\langle 0|^c\otimes
G'_1\otimes G'_{2,...,n}+|1\rangle\langle 1|^c\otimes
OG'_1O^{\dagger}\otimes G'_{2,...,n}$ which can be used (with a
minus sign) as a starting Hamiltonian for a subsequent operation. In
particular, we can implement the conditional $Q$ following the
conditional $O$ using the Hamiltonian
\begin{equation}
\widehat{\widehat{H}}_{C(Q)}(t)=-|0\rangle\langle 0|^c\otimes
G'_1\otimes G'_{2,...,n}-\beta(t)|1\rangle\langle 1|^c\otimes
H_Q(t)\otimes G'_{2,...,n}, \label{HamB}
\end{equation}
where the factor $\beta(t)$ guarantees that there is no splitting of
the energy levels. Subsequent operations can be applied analogously.
Using this general method, we can implement a unitary whose
geometric part is equal to any transversal Clifford operation
conditioned on the ``cat" state.

\subsection{Preparing and using the ``cat" state}

In addition to transversal operations, a complete fault-tolerant
scheme requires the ability to prepare, verify and use a special
ancillary state such as the ``cat" state
$(|00...0\rangle+|11...1\rangle)/\sqrt{2}$ proposed by Shor
\cite{Shor96}. This can also be done using our geometric approach.
Since the ``cat" state is known and its construction is
non-fault-tolerant, we can prepare it by simply treating each
initially prepared qubit as a simple code (with $\widetilde{G}$ in
Eq.~\eqref{stabelement} being trivial), and updating the stabilizer
of the code via the applied geometric transformation as the
operation progresses. The stabilizer of the prepared ``cat" state is
generated by $Z_iZ_j$, $i<j$. Transversal unitary operations between
the ``cat" state and other codewords are applied as described in the
previous section.

We also have to be able to measure the parity of the state, which
requires the ability to apply successive C-NOT operations from two
different qubits in the ``cat" state to the same ancillary qubit
initially prepared in the state $|0\rangle$. We
can regard a qubit in state $|0\rangle$ as a simple code with stabilizer $%
\langle Z \rangle $, and we can apply the first C-NOT as described
before. Even though after this operation the state of the target
qubit is unknown, the second C-NOT gate can be applied via the same
interaction, since the transformation undergone by each eigenspace
would still be equivalent to the desired C-NOT and at the end when
we measure the qubit we project onto one of the eigenspaces.

\subsection{Fault tolerance of the scheme}

We showed how we can generate any transversal operation on the code
space geometrically, assuming that the state is non-erroneous. But
what if an error occurs on one of the qubits?

At any moment, we can distinguish two types of errors---those that
result in transitions between the ground and the excited spaces of
the current Hamiltonian, and those that result in transformations
inside the eigenspaces. Due to the discretization of errors in QEC,
it suffices to prove correctability for each type separately. The
key property of our construction is that the transformation
undergone by each of the eigenspaces is equivalent to the same
transversal operation. Because of this, if we are applying a unitary
on the first qubit, an error on that qubit will remain localized
regardless of whether it causes an excitation or not. If the error
occurs on one of the other qubits, at the end of the transformation
the result would be the desired single-qubit unitary gate plus the
error on the other qubit, which is correctable.

It is remarkable that even though the Hamiltonian couples qubits
within the same block, single-qubit errors do not propagate. This is
because the coupling between the qubits amounts to a change in the
relative phase between the ground and excited spaces, but the latter
is irrelevant since either it is equivalent to a gauge
transformation, or when we apply a correcting operation we project
onto one of the eigenspaces. In the case of the C-NOT gate, an error
can propagate between the control and the target qubits, but it
never results in two errors within the same codeword.

\section{Effects on the accuracy threshold for environmental noise}

Since the method we presented conforms completely to a given
fault-tolerant scheme, it would not affect the error threshold per
operation for that scheme. Some of its features, however, would
affect the threshold for \textit{environment} noise.

First, observe that when applying the Hamiltonian \eqref{Ham1}, we
cannot at the same time apply
operations on the other qubits on which the factor $%
\widetilde{G}$ acts non-trivially. Thus, some operations at the
lowest level of concatenation that would otherwise be implemented
simultaneously might have to be implemented serially. The effect of
this is equivalent to slowing down the circuit by a constant factor.
(Note that we could also vary the factor $\widetilde{G}$
simultaneously with $H(t)$, but in order to obtain the same
precision as that we would achieve by a serial implementation, we
would have to slow down the change of the Hamiltonian by the same
factor.) The slowdown factor resulting from this loss of parallelism
is usually small since this problem occurs only at the lowest level
of concatenation. For example, when implementing encoded
single-qubit gates with the Bacon-Shor code, we can apply operations
on up to 6 out of the 9 qubits in a block simultaneously. As we show
in Sec. V, we can address any two qubits in a row or column using
our method by taking $\widetilde{G}$ in Eq.~~\eqref{Ham1} to be a
single-qubit operator $Z$ or $X$ on the third qubit in the same row
or column. The Hamiltonians used to apply operations on the two
qubits commute with each other at all times and do not interfere. A
similar phenomenon holds for the implementation of the encoded C-NOT
gate, or the operations involving the ``cat" state. Thus, for the
Bacon-Shor code we have a slowdown due to parallelism by a factor of
$1.5$.

A more significant slowdown results from the fact that the evolution
is adiabatic. In order to obtain a rough estimate of the slowdown
due specifically to the adiabatic requirement, we will compare the
time $T_{h}$ needed for the adiabatic implementation of a given gate
with precision $1-\delta$ to the time $T_{d}$ needed for a dynamical
realization of the same gate with the same strength of the
Hamiltonian. We will consider a realization of the $X$ gate via the
unitary interpolation \cite{Siu}
\begin{equation}
\widehat{H}(t)=-V_{X}(\tau(t))ZV_{X}^{\dagger}(\tau(t))\otimes
\widetilde{G}, \hspace{0.2cm} V_{X}(\tau(t))=\textrm{exp}
\left(i\tau(t)\frac{\pi }{2T_{h}}X\right),\label{Hamest}
\end{equation}
where $\tau(0)=0$, $\tau(T_h)=T_h$. Thus the energy gap of the
Hamiltonian is constant. The optimal dynamical implementation of the
same gate is via the Hamiltonian $-X$ for time
$T_{d}=\frac{\pi}{2}$.

As we argued in Sec. III, the accuracy with which the adiabatic
approximation holds for the Hamiltonian \eqref{Hamest} is the same
as that for the Hamiltonian
\begin{equation}
H(t)=V_{X}(\tau(t))ZV_{X}^{\dagger }(\tau(t)).\label{Hamest2}
\end{equation}
We now present estimates for two different choices of the function
$\tau(t)$. The first one is
\begin{equation}
\tau(t)=t.
\end{equation}
In this case the Schr\"{o}dinger equation can be easily solved in
the instantaneous eigenbasis of the Hamiltonian \eqref{Hamest2}.
For the probability that the initial ground state remains a ground
state at the end of the evolution, we obtain
\begin{equation}
p=\frac{1}{1+\varepsilon ^{2}}+\frac{%
\varepsilon ^{2}}{1+\varepsilon ^{2}}\cos ^{2}(\frac{\pi }{4\varepsilon }%
\sqrt{1+\varepsilon ^{2}})=1-\delta,
\end{equation}
where
\begin{equation}
\varepsilon =\frac{T_{d}}{T_{h}}.
\end{equation}
Expanding in powers of $\varepsilon $ and averaging the square of
the cosine whose period is much smaller than $T_{h}$, we obtain
the condition
\begin{equation}
\varepsilon ^{2}\leq 2\delta.
\end{equation}
Assuming, for example, that $\delta \approx 10^{-4}$ (approximately
the threshold for the 9-qubit Bacon-Shor code \cite{AC07}), we obtain
that the time of evolution for the adiabatic case must be about 70
times longer than that in the dynamical case.

It is known, however, that if $H(t)$ is smooth and all its
derivatives vanish at $t=0$ and $t=T_h$, the adiabatic error
decreases super-polynomially with $T_h$ \cite{HJ02}. To achieve
this, we will choose
\begin{equation}
\tau(t)= \frac{1}{a}\int_0^t dt' e^{-1/\sin(\pi
t'/T_h)},\hspace{0.2cm} a=\int_0^{T_h} dt' e^{-1/\sin(\pi
t'/T_h)}.
\end{equation}
For this interpolation, by a numerical solution we obtain that
when $T_h/T_d\approx 17$ the error is already of the order of
$10^{-6}$, which is well below the threshold values obtained for
the Bacon-Shor codes \cite{AC07}. This is a remarkable improvement
in comparison to the previous interpolation which shows that the
smoothness of the Hamiltonian plays an important role in the
performance of the scheme.

An additional slowdown in comparison to a perfect dynamical scheme
may result from the fact that the constructions for some of the
standard gates we presented involve long sequences of loops. With
more efficient parameter paths, however, it should be possible to
reduce this slowdown to minimum. An approach for finding optimal
loops presented in Ref.~\cite{TNH04} may be useful in this respect.

In comparison to a dynamical implementation, the allowed rate of
environmental noise for the holonomic case would decrease by a
factor similar to the slowdown factor. In practice, however,
dynamical gates are not perfect and the holonomic approach may be
advantageous if it gives rise to a better operational precision.

We finally point out that an error in the factor $H(t)$ in the
Hamiltonian \eqref{Ham1} would result in an error on the first qubit
according to Eq.~\eqref{finalU}. Such an error clearly has to be
below the accuracy threshold. More dangerous errors, however, are
also possible. For example, if the degeneracy of the Hamiltonian is
broken, this can result in an unwanted dynamical transformation
affecting all qubits on which the Hamiltonian acts non-trivially.
Such multi-qubit errors have to be of higher order in the threshold,
which imposes more severe restrictions on the Hamiltonian.

\section{Fault-tolerant holonomic computation with low-weight
Hamiltonians}

The weight of the Hamiltonians needed for the scheme we described
depends on the weight of the stabilizer or gauge-group elements.
Remarkably, certain codes possess stabilizer or gauge-group elements
of low weight covering all qubits in the code, which allows us to
perform holonomic computation using low-weight Hamiltonians. Here we
will consider as an example a subsystem generalization of the
9-qubit Shor code \cite{Shor95}---the Bacon-Shor code \cite{Bac06,
BC06}---which has particularly favorable properties for
fault-tolerant computation \cite{AC07, Ali07}. In the 9-qubit
Bacon-Shor code, the gauge group is generated by the weight-two
operators $Z_{k,j}Z_{k,j+1}$ and $X_{j,k}X_{j+1,k}$, where the
subscripts label the qubits by row and column when they are arranged
in a $3\times 3$ square lattice. Since the Bacon-Shor code is a CSS
code, the C-NOT gate has a direct transversal implementation. We now
show that the C-NOT gate can be realized using at most weight-three
Hamiltonians.

If we want to apply a C-NOT gate between two qubits each of which
is, say, in the first row and column of its block, we can use as a
starting Hamiltonian $-Z^{t}_{1,1}\otimes Z^t_{1,2}$, where the
superscript $t$ signifies that these are operators in the target
block. We can then apply the C-NOT gate as described in Sec. III.
After the operation, however, this gauge-group element will
transform to $-Z^{t}_{1,1}\otimes Z^c_{1,1}\otimes Z^t_{1,2}$. If we
now want to implement a C-NOT gate between the qubits with index
$\{1,2\}$ using as a starting Hamiltonian the operator
$-Z^{t}_{1,1}\otimes Z^c_{1,1}\otimes Z^t_{1,2}$ according to the
same procedure, we will have to use a four-qubit Hamiltonian. Of
course, at this point we can use the starting Hamiltonian
$-Z^t_{1,2}\otimes Z^t_{1,3}$, but if we had also applied a C-NOT
between the qubits labeled $\{1,3\}$, this operator would not be
available---it would have transformed to $-Z^t_{1,2}\otimes
Z^t_{1,3}\otimes Z^c_{1,3}$.

What we can do instead, is to use as a starting Hamiltonian the
operator $ -Z^{t}_{1,1}\otimes Z^t_{1,2}\otimes Z^c_{1,2}$ which
is obtained from the gauge-group element $ Z^{t}_{1,1}\otimes
Z^c_{1,1}\otimes Z^t_{1,2}\otimes Z^c_{1,2}$ after the application
of the C-NOT between the qubits with index $\{1,1\}$. Since the
C-NOT gate is its own inverse, we can regard the factor
$Z^{t}_{1,1}$ as $\widetilde{G}$ in Eq.~\eqref{CNOTint} and use
this starting Hamiltonian to apply our procedure backwards. Thus
we can implement any transversal C-NOT gate using at most
weight-three Hamiltonians.

Since the encoded $X$, $Y$ and $Z$ operations have a bitwise
implementation, we can always apply them according to our procedure
using Hamiltonians of weight 2. For the Bacon-Shor code, the encoded
Hadamard gate can be applied via bitwise Hadamard transformations
followed by a rotation of the grid by a $90$ degree angle
\cite{AC07}. The encoded Phase gate can be implemented by using the
encoded C-NOT and an ancilla.

The preparation and measurement of the ``cat" state can also be done
using Hamiltonians of weight 2. To prepare the ``cat" state, we
first prepare all qubits in the state $(|0\rangle +
|1\rangle)/\sqrt{2}$, which can be done by measuring each of them in
the $\{|0\rangle,|1\rangle\}$ basis (this ability is assumed for any
type of computation) and applying the transformation $-Z\rightarrow
-X$ or $Z\rightarrow -X$ depending on the outcome. To complete the
preparation of the ``cat" state, apply a two-qubit transformation
between the first qubit and each of the other qubits ($j>1$) via the
transformation
\begin{equation}
-I_1\otimes X_j\rightarrow -Z_1\otimes Z_j .
\end{equation}
Single-qubit transformations on qubits from the ``cat" state can
be applied according to the method described in the previous
section using at most weight-two Hamiltonians.

To measure the parity of the state, we need to apply successively
C-NOT operations from two different qubits in the ``cat" state to
the same ancillary qubit initially prepared in the state
$|0\rangle$. As described in Sec. III, this can also be done
according to our method and requires Hamiltonians of weight 2.

For universal computation with the Bacon-Shor code, we also need to
be able to apply one encoded transformation outside of the Clifford
group. As we mentioned earlier, in order to implement the Toffoli
gate or the $\pi/8$ gate, it is sufficient to be able to implement a
C-NOT gate conditioned on a ``cat" state. For the Bacon-Shor code,
the C-NOT gate has a transversal implementation, so the conditioned
C-NOT gate can be realized by a series of transversal Toffoli
operations between the ``cat" state and the two encoded states. We
now show that this gate can be implemented using at most three-qubit
Hamiltonians.

Ref.~\cite{NieChu00} provides a circuit for implementing the
Toffoli gate as a sequence of one- and two-qubit gates. We will
use the same circuit, except that we flip the control and target
qubits in every C-NOT gate using the identity
\begin{equation}
(R_{1}\otimes R_{2}) C_{1,2}(R_{1}\otimes R_{2})= C_{2,1},
\end{equation}
where $R_{i}$ denotes a Hadamard gate on the qubit labeled by $i$
and $C_{i,j}$ denotes a C-NOT gate between qubits $i$ and $j$ with
$i$ being the control and $j$ being the target. Let
$\textrm{Toffoli}_{i,j,k}$ denote the Toffoli gate on qubits $i$,
$j$ and $k$ with $i$ and $j$ being the two control qubits and $k$
being the target qubit, and let $S_{i}$ and $T_i$ denote the Phase
and $\pi/8$ gates on qubit $i$, respectively. Then the Toffoli gate
on three qubits (the first one of which we will assume to belong to
the ``cat" state), can be written as:
\begin{gather}
\textrm{Toffoli}_{1,2,3}=R_{2}C_{3,2}R_{3}T_{3}^{\dagger}R_{3}R_{1}C_{3,1}R_{3}T_{3}R_{3}C_{3,2}R_{3}T_{3}^{\dagger}R_{3}
C_{3,1}R_{3}T_{3}R_{3}R_{2}T_{2}^{\dagger}R_{2}
C_{2,1}R_{2}T_{2}^{\dagger}R_{2}
C_{2,1}R_{2}S_{2}R_{1}T_{1}.\label{Toffoli}
\end{gather}
To show that each of the above gates can be implemented according to
our geometric approach using Hamiltonians of weight at most 3, we
will need an implementation of the C-NOT gate which is suitable for
the case when we have a stabilizer or gauge-group element of the
form
\begin{equation}
\widehat{G}=X\otimes \widetilde{G},\label{Xgenerator}
\end{equation}
where the factor $X$ acts on the target qubit and $\widetilde{G}$
acts trivially on the control qubit. By a similar argument to the
one in Sec. III, one can verify that in this case the C-NOT gate can
be implemented as follows: apply the operation $S^{\dagger}$ on the
control qubit (we describe how to do this for our particular case
below) together with the transformation
\begin{equation}
-X\otimes\widetilde{G}\rightarrow - Z\otimes
\widetilde{G}\rightarrow X\otimes\widetilde{G}\label{CNOT1}
\end{equation}
on the target qubit, followed by the transformation
\begin{equation}
I^c\otimes X\otimes \widetilde{G}\rightarrow -(|0\rangle\langle
0|^c\otimes Z+|1\rangle\langle 1|^c\otimes Y)\otimes
\widetilde{G}\rightarrow-I^c\otimes X\otimes
\widetilde{G}.\label{CNOT2}
\end{equation}

Since the second and the third qubits belong to blocks encoded with
the Bacon-Shor code, there are weight-two elements of the initial
gauge group of the form $Z\otimes Z$ covering all qubits. The
stabilizer generators on the ``cat" state are also of this type.
Following the transformation of these operators according to the
sequence of operations \eqref{Toffoli}, one can see that before
every C-NOT gate in this sequence, there is an element of the form
\eqref{Xgenerator} with $\widetilde{G}=Z$ that can be used to
implement the C-NOT gate as described, provided that we can
implement the gate $S^{\dagger}$ on the control qubit. We also point
out that all single-qubit operations on qubit $1$ in this sequence
can be implemented according to the procedure described in Sec. III,
since at every step we have a weight-two stabilizer element on that
qubit with a suitable form. Therefore, all we need to show is how to
implement the necessary single-qubit operations on qubits $2$ and
$3$. Due to the complicated transformation of the gauge-group
elements during the sequence of operations \eqref{Toffoli}, we
introduce a method of applying a single-qubit operation with a
starting Hamiltonian that acts trivially on the qubit. For
implementing single-qubit operations on qubits $2$ and $3$ we
use as a starting Hamiltonian the operator
\begin{equation}
\widehat{\widehat{H}}(0)=-I_{i}\otimes X_1\otimes \widetilde{Z},
\hspace{0.4 cm} i=2,3,\label{specialcase}
\end{equation}
where the first factor ($I_i$) acts on the qubit on which we want
to apply the operation ($2$ or $3$), and $X_1\otimes
\widetilde{Z}$ is the transformed (after the Hadamard gate $R_1$)
stabilizer element of the ``cat" state that acts non-trivially on
qubit $1$ (the factor $\widetilde{Z}$ acts on some other qubit in
the ``cat" state).

To implement a single-qubit gate on qubit $3$ for example, we
first apply the interpolation
\begin{equation}
-I_{3}\otimes X_1\otimes \widetilde{Z}\rightarrow -Z_{3}\otimes
Z_1\otimes \widetilde{Z}.\label{nexttolast}
\end{equation}
This results in a two-qubit geometric transformation $U_{1,3}$ on
qubits $1$ and $3$. We do not have to calculate this transformation
exactly since we will undo it later, but the fact that each
eigenspace undergoes the same two-qubit geometric transformation can
be verified similarly to the C-NOT gate we described in Sec. III.

At this point, the Hamiltonian is of the form \eqref{H(0)} with
respect to qubit 3, and we can apply any single-qubit unitary gate
$V_3$ according to the method described in Sec. III. This transforms
the Hamiltonian to $-V_3Z_{3}V_3^{\dagger}\otimes Z_1\otimes
\widetilde{Z}$. We can now ``undo" the transformation $U_{1,3}$ by
the interpolation
\begin{equation}
-V_3Z_{3}V_{3}^{\dagger}\otimes Z_1\otimes
\widetilde{Z}\rightarrow-I_{3}\otimes X_1\otimes
\widetilde{Z}.\label{lastHamiltonian}
\end{equation}
The latter transformation is the inverse of Eq.~\eqref{nexttolast}
up to the single-qubit unitary transformation $V_3$, i.e., it
results in the transformation $V_3U_{1,3}^{\dagger}V_{3}^{\dagger}$.
Thus the net result is
\begin{equation}
V_3U_{1,3}^{\dagger}V_3^{\dagger}V_3U_{1,3}=V_3,
\end{equation}
which is the desired single-qubit unitary transformation on qubit
$3$. We point out that during this transformation, a single-qubit
error can propagate between qubits $1$ and $3$, but this is not a
problem since we are implementing a transversal Toffoli operation
and such an error would not result in more than one error per block
of the code.

We showed that for the Bacon-Shor code our scheme can be implemented
with at most 3-local Hamiltonians. This is optimal for the
construction we presented, since there are no non-trivial codes with
stabilizer or gauge-group elements of weight smaller than 2 covering
all qubits. One could argue that since the only Hamiltonians that
leave the code space invariant are superpositions of elements of the
stabilizer or the gauge group, one cannot do better than this.
However, it may be possible to approximate the necessary
Hamiltonians with sufficient precision using 2-local interactions. A
possible direction to consider in this respect are the gadget
techniques introduced in Ref.~\cite{KKR06} and developed further in
Refs.~\cite{OT05, Jordan08}. This is left as a problem for future
investigation.

\section{Conclusion}

We described a scheme for fault-tolerant holonomic computation on
stabilizer codes, which demonstrates that HQC is a scalable method
of computation. The scheme opens the possibility for combining the
software protection of error correction with the inherent robustness
of HQC against control imperfections. Our construction uses
Hamiltonians that are elements of the stabilizer or the gauge group
of the code and works by adiabatically varying these Hamiltonians in
a manner which generates unitaries whose geometric parts are equal
to transversal operations from the point of view of the basis of the
full Hilbert space. By composing these transversal operations as in
a given standard dynamical fault-tolerant scheme, we thus transport
the code space (or the redundant subspaces that constitute a code
subsystem) adiabatically around loops that give rise to holonomies
equal to encoded operations inside the code space. The Hamiltonians
needed for implementing two-qubit gates are at least 3-local. We
showed that computation with at most 3-local Hamiltonians is
possible with the Bacon-Shor code.

It is interesting to point out that the adiabatic regime in which
our scheme operates is consistent with the model of Markovian
decoherence. In Ref.~\cite{ALZ06} it was argued that the standard
dynamical paradigm of fault tolerance is based on assumptions that
are in conflict with the rigorous derivation of the Markovian limit.
Although the threshold theorem has been extended to non-Markovian
models \cite{TB05, AGP06, AKP06}, the Markovian assumption is an
accurate approximation for a wide range of physical scenarios
\cite{Car99}. It also allows for a much simpler description of the
evolution in comparison to non-Markovian models (see, e.g., Ref.~
\cite{OB07}). In Ref.~\cite{ALZ06} it was shown that the
weak-coupling-limit derivation of the Markovian approximation is
consistent with computational methods that employ slow
transformations, such as adiabatic quantum computation \cite{FGGS00}
or HQC. A theory of fault tolerance for the adiabatic model of
computation at present is not known, although some steps in this
direction have been undertaken \cite{JFS06, Lid07}. Our hybrid
HQC-QEC scheme provides a solution for the case of HQC. We point
out, however, that it is an open problem whether the Markovian
approximation makes sense for a fixed value of the adiabatic
slowness parameter when the circuit increases in size. Giving a
definitive answer to this question requires a rigorous analysis of
the accumulation of non-Markovian errors due to deviation from
perfect adiabaticity.

Applying the present strategy to actual physical systems might
require modifying our abstract construction in accordance with the
available interactions, possibly using linear combinations of
stabilizer or gauge-group elements rather than single elements as
the basic Hamiltonians. Given that simple QEC codes and two-qubit
geometric transformations have been realized using NMR \cite{NMR1, NMR2} and ion-trap \cite%
{Tra1, Tra2} techniques, these systems seem particularly suitable
for hybrid HQC-QEC implementations.

We hope that the techniques presented in this paper might prove
useful in other areas as well. It is possible that some combination
of transversal adiabatic transformations and active correction could
provide a solution to the problem of fault tolerance in the
adiabatic model of computation as well.

\section*{Acknowledgements} OO\ acknowledges support under NSF Grant No. CCF-0524822 and
Spanish MICINN (Consolider-Ingenio QOIT); TAB acknowledges support
under NSF Grant No. CCF-0448658; DAL acknowledges support under NSF
Grant No. PHY-803304 and NSF Grant No. CCF-0726439.

DAL gratefully acknowledges the hospitality of the Caltech Institute
for Quantum Information (IQI), where part of this work was
performed. IQI is supported in part by the National Science
Foundation under Grant No. PHY-0803371.

\appendix

\section{Proof of Proposition 2}

To prove Proposition 2, observe that the Hermitian unitary matrix
\begin{equation}
W^{\theta}=\begin{pmatrix} 0&ie^{-i\theta}\\-ie^{i\theta}&0
\end{pmatrix}\label{1}
\end{equation}
has the properties
\begin{equation}
[W^{\theta},V^{\theta\pm}]=0,
\end{equation}
\begin{equation}
\{W^{\theta},Z\}=0,
\end{equation}
where $[\cdot,\cdot]$ and $\{\cdot,\cdot\}$ denote a commutator and
an anticommutator, respectively. This means that
\begin{equation}
W^{\theta}H^{\theta\pm}(t)W^{\theta}=
W^{\theta}(f(t)Z+g(t)V^{\theta\pm}Z
V^{\theta\mp})W^{\theta}=-f(t)Z-g(t)V^{\theta\pm}Z
V^{\theta\mp}=-H^{\theta\pm}(t),
\end{equation}
i.e., the unitary $W^{\theta}$ flips the ground and excited spaces
of $H^{\theta\pm}(t)$ for any $t$.

The unitaries $U^{\theta\pm}_{0,1}$, given by Eq.~\eqref{unitaries}
for $H(t)=H^{\theta\pm}(t)$, are therefore related by
\begin{equation}
U^{\theta\pm}_0(t)=W^{\theta} U^{\theta\pm}_1(t)
W^{\theta}.\label{U01}
\end{equation}
Using the fact that
\begin{gather}
W^{\theta}|0\rangle=-ie^{i\theta}|1\rangle,\notag\\
W^{\theta}|1\rangle=ie^{-i\theta}|0\rangle,\label{12}
\end{gather}
from Eq.~\eqref{unita} and Eq.~\eqref{holon} one can see that
Eq.~\eqref{U01} implies
\begin{equation}
U^{\theta\pm}_{A_0}(t)=W^{\theta}U^{\theta\pm}_{A_1}(t)W^{\theta}.
\label{wrelation}
\end{equation}
Let us define the eigenstates of $H^{\theta\pm}(t)$ at time $T$ as
$|\phi^{\theta\pm}_0(T)\rangle = V^{\theta\pm}|0\rangle$ and
$|\phi^{\theta\pm}_1(T)\rangle = V^{\theta\pm}|1\rangle$. Expression
\eqref{holon} can then be written as
\begin{gather}
U^{\theta\pm}_{A_{0}}(T)=e^{i\alpha^{\theta\pm}_0}
V^{\theta\pm}|0\rangle \langle 0|,\notag\\
U^{\theta\pm}_{A_{1}}(T)=e^{i\alpha^{\theta\pm}_1}
V^{\theta\pm}|1\rangle \langle 1|,\label{UA01}
\end{gather}
where $\alpha^{\theta\pm}_0$ and $\alpha^{\theta\pm}_1$ are
geometric phases. Without explicitly calculating the geometric
phases, from Eq.~\eqref{UA01}, Eq.~\eqref{wrelation}, Eq.~\eqref{12}
and Eq.~\eqref{1}, we obtain
\begin{gather}
e^{i\alpha^{\theta\pm}_0} V^{\theta\pm}|0\rangle \langle
0|=e^{i\alpha^{\theta\pm}_1} W^{\theta}V^{\theta\pm}|1\rangle
\langle 1|W^{\theta}=e^{i\alpha^{\theta\pm}_1}
V^{\theta\pm}W^{\theta}|1\rangle \langle
1|W^{\theta}=e^{i\alpha^{\theta\pm}_1} V^{\theta\pm}|0\rangle
\langle 0|,
\end{gather}
i.e.,
\begin{equation}
e^{i\alpha^{\theta\pm}_0}=e^{i\alpha^{\theta\pm}_1}.
\end{equation}
Therefore, up to a global phase, Eq.~\eqref{finalU} yields
\begin{equation}
U^{\theta\pm}(T)\sim V^{\theta\pm}.
\end{equation}

\section{Calculating the holonomy for the $Z$ gate}
\label{AppendixB}

\subsection{Linear interpolation}

We first demonstrate how to calculate the ground-space holonomy for
the $Z$ gate for the case of linear interpolations along the
segments of the path, i.e., when $f(t)$ and $g(t)$ in
Eq.~\eqref{interpolation} are
\begin{equation}
f(t)=1-\frac{t}{T}, \hspace{0.4cm}g(t)=\frac{t}{T}.
\end{equation}

In order to calculate the holonomy \eqref{holonomy} corresponding to
our construction of the $Z$ gate, we need to define a
\textit{single-valued} orthonormal basis of the ground space of the
Hamiltonian along the loop described by Eq.~\eqref{Zgate}. Since the
Hamiltonian has the form \eqref{Ham2} at all times, it is convenient
to choose the basis of the form
\begin{gather}
|j k; \lambda\rangle =
|\chi_j(\lambda)\rangle|\widetilde\psi_{j k}\rangle,\\
\hspace{0.2cm} j=0,1; \hspace{0.2cm}
k=1,...,2^{n-2}\notag\hspace{0.2cm},
\end{gather}
where $|\chi_0(\lambda(t))\rangle$ and $|\chi_1(\lambda(t))\rangle$
are ground and excited states of $H(t)$, and
$|\widetilde\psi_{0k}\rangle$ and $|\widetilde\psi_{1k}\rangle$ are
fixed orthonormal bases of the subspaces that support the projectors
$\widetilde{P}_0$ and $\widetilde{P}_1$ defined in
Eq.~\eqref{Projectors}, respectively. The eigenstates
$|\chi_0(\lambda(t))\rangle$ and $|\chi_1(\lambda(t))\rangle$ are
defined up to an overall phase, but we have to choose the phase such
that the states are single-valued along the loop.

Observe that because of this choice of basis, the matrix elements
\eqref{matrixelementsA} become
\begin{gather}
({A_\mu})_{jk, j'k'}=\langle jk; \lambda| \frac{\partial}{\partial
\lambda^{\mu}}|j'k'; \lambda\rangle=\langle
\chi_j(\lambda)|\frac{\partial}{\partial\lambda^\mu}|\chi_{j'}(\lambda)\rangle\notag\\
\times\langle\widetilde{\psi}_{jk}|\widetilde{\psi}_{j'k'}\rangle
=\langle
\chi_j(\lambda)|\frac{\partial}{\partial\lambda^\mu}|\chi_{j'}(\lambda)\rangle
\delta_{jj'}\delta_{kk'},
\end{gather}
i.e., the matrix $A_\mu$ is diagonal. (Since we are looking only at
the ground space, we are not writing the index of the energy level.)
We can therefore drop the path-ordering operator. The resulting
unitary matrix $U^{\gamma}_{jk,j'k'}$ acting on the subspace spanned
by $\{|jk;\lambda(0)\rangle\}$ is also diagonal and its diagonal
elements are
\begin{equation}
U^{\gamma}_{jk,jk}=\textrm{exp}\left(\oint_{\gamma}\langle
\chi_j(\lambda)|\frac{\partial}{\partial\lambda^\mu}|\chi_{j}(\lambda)\rangle
d\lambda^\mu\right).
\end{equation}
These are precisely the Berry phases for the loops described by the
states $|\chi_{j}(\lambda\rangle)$. Since the loop in parameter
space consists of four line segments, we can write the last
expression as
\begin{equation}
U^{\gamma}_{jk,jk}=\textrm{exp}\left(\sum_{i=1}^4 \int_{\gamma_i}
\langle
\chi_j(\lambda)|\frac{\partial}{\partial\lambda^\mu}|\chi_{j}(\lambda)\rangle
d\lambda^\mu\right),
\end{equation}
where $\gamma_i$, $i=1,2,3,4$, are the segments indexed by their
order corresponding to Eq.~\eqref{Zgate}. If we parametrize each
line segment by the dimensionless time $0\leq s \leq 1$, we obtain
\begin{equation}
U^{\gamma}_{jk,jk}=\textrm{exp}\left(\sum_{i=1}^4 \int_0^1 \langle
\chi_j^i(s)|\frac{d}{ds}|\chi_{j}^i(s)\rangle
ds\right),\label{diagonalU}
\end{equation}
where the superscript $i$ in $|\chi_{j}^i(s)\rangle$ indicates the
segment. In the $\{|0\rangle, |1\rangle\}$ basis, we will write
these states as
\begin{equation}
|\chi_j^i(s)\rangle=\begin{pmatrix}
a^i_j(s)\\
b^i_j(s)
\end{pmatrix}, \hspace{0.2cm}
j=0,1\hspace{0.2cm},\hspace{0.2cm}i=1,2,3,4\hspace{0.2cm},
\end{equation}
where $|a^i_j(s)|^2+|b^i_j(s)|^2=1$.

Along the segment $\gamma_1$, the Hamiltonian has the form
$\widehat{H}(s)=H_1(s)\otimes\widetilde{P}_0-H_1(s)\otimes\widetilde{P}_1$,
where
\begin{equation}
H_1(t)=(1-s)Z+sX,
\end{equation}
i.e, the states $|\chi_0^1(s)\rangle$ and $|\chi_1^1(s)\rangle$ are
the ground and excited states of $H_1(s)$. For these states we
obtain
\begin{gather}
a^1_0(s)=\frac{(1-s+\sqrt{1-2s+2s^2})e^{i\omega^1_0(s)}}{\sqrt{2-4s+4s^2+(2-2s)\sqrt{1-2s+2s^2}}},\label{a30}\\
b^1_0(s)=\frac{s e^{i\omega^1_0(s)}}{\sqrt{2-4s+4s^2+(2-2s)\sqrt{1-2s+2s^2}}},\label{b30}\\
a^1_1(s)=\frac{(1-s-\sqrt{1-2s+2s^2})e^{i\omega^1_1(s)}}{\sqrt{2-4s+4s^2-(2-2s)\sqrt{1-2s+2s^2}}},\label{a31}\\
b^1_1(s)=\frac{se^{i\omega^1_1(s)}}{\sqrt{2-4s+4s^2-(2-2s)\sqrt{1-2s+2s^2}}},\label{b31}
\end{gather}
where $\omega^1_j(s)$ are arbitrary phases which have to be chosen
so that when we complete the loop, the phases of the corresponding
states will return to their initial values modulo $2\pi$. We will
define the loops as interpolating between the following
intermediate states defined with their overall phases:
\begin{gather}
|\psi_0(\lambda)\rangle: \hspace{0.2cm}|0\rangle \rightarrow
|f_+^0\rangle \rightarrow |1\rangle \rightarrow
|f_-^{\pi/2}\rangle \rightarrow |0\rangle,\\
|\psi_1(\lambda)\rangle: \hspace{0.2cm}|1\rangle \rightarrow
|f_-^0\rangle \rightarrow |0\rangle \rightarrow
|f_+^{\pi/2}\rangle \rightarrow |1\rangle,
\end{gather}
where
\begin{equation}
|f_{\pm}^\theta\rangle =\frac{|0\rangle \pm
e^{i\theta}|1\rangle}{\sqrt{2}}.
\end{equation}
In other words, we impose the conditions
$|\chi_{0,1}^1(0)\rangle=|0,1\rangle$,
$|\chi_{0,1}^1(1)\rangle=|f_{\pm}^0\rangle =
|\chi_{0,1}^2(0)\rangle$, $|\chi_{0,1}^2(1)\rangle =|1,0\rangle
=|\chi_{0,1}^3(0)\rangle$,
$|\chi_{0,1}^3(1)\rangle=|f_{\mp}^{\pi/2}\rangle=|\chi_{0,1}^4(0)\rangle$,
$|\chi_{0,1}^4(1)\rangle=|0,1\rangle$.

From Eq.~\eqref{a30} and Eq.~\eqref{b30} we see that
$a^1_0(0)=e^{i\omega^1_0(0)}$, $b^1_0(0)=0 $ and
$a^1_0(1)=\frac{1}{\sqrt{2}}e^{i\omega^1_0(1)}$,
$b^1_0(1)=\frac{1}{\sqrt{2}}e^{i\omega^1_0(1)}$ , so we can choose
\begin{equation}
\omega^1_0(s)=0,\hspace{0.2cm} \forall s\in [0,1].
\end{equation}
Similarly, from Eq.~\eqref{a31} and Eq.~\eqref{b31} it can be seen
that $a^1_1(0)=0$, $b^1_1(0)=e^{i\omega^1_1(0)} $ and
$a^1_1(1)=-\frac{1}{\sqrt{2}}e^{i\omega^1_1(1)}$,
$b^1_1(1)=\frac{1}{\sqrt{2}}e^{i\omega^1_1(1)}$. This means that
$\omega^1_1(s)$ has to satisfy $e^{i\omega^1_1(0)}=1$,
$e^{i\omega^1_1(1)}=-1$. We can choose any differentiable
$\omega^1_1(s)$ that satisfies
\begin{equation}
\omega^1_1(0)=0, \hspace{0.2cm} \omega^1_1(1)=\pi.
\end{equation}

In order to calculate $\int_0^1 \langle
\chi_j^1(s)|\frac{d}{ds}|\chi_{j}^1(s)\rangle ds$, we also need
\begin{equation}
\frac{d}{ds}|\chi_{j}^1(s)\rangle=
\begin{pmatrix} \frac{d}{ds}a^1_j(s)\\
\frac{d}{ds}b^1_j(s)
\end{pmatrix}.
\end{equation}
Differentiating Eqs.~\eqref{a30}-\eqref{b31} yields
\begin{widetext}
\begin{gather}
\frac{d}{ds}a^1_0(s)=-\frac{s(1-s+\sqrt{1-2s+2s^2})}{2\sqrt{2-4s+4s^2}[1-2s+2s^2+(1-s)\sqrt{1-2s+2s^2}]^{\frac{3}{2}}},\\
\frac{d}{ds}b^1_0(s)=\frac{2-4s+3s^2+(2-2s)\sqrt{1-2s+2s^2}}{2\sqrt{2-4s+4s^2}[1-2s+2s^2+(1-s)\sqrt{1-2s+2s^2}]^{\frac{3}{2}}},\\
\frac{d}{ds}a^1_1(s)=-\frac{s(1-s-\sqrt{1-2s+2s^2})e^{i\omega^1_1(s)}}{2\sqrt{2-4s+4s^2}[1-2s+2s^2-(1-s)\sqrt{1-2s+2s^2}]^{\frac{3}{2}}} +a^1_1(s)i\frac{d}{ds}\omega^1_1(s),\\
\frac{d}{ds}b^1_0(s)=-\frac{(2-4s+3s^2-(2-2s)\sqrt{1-2s+2s^2})e^{i\omega^1_1(s)}}{2\sqrt{2-4s+4s^2}[1-2s+2s^2-(1-s)\sqrt{1-2s+2s^2}]^{\frac{3}{2}}}
+b^1_1(s)i\frac{d}{ds}\omega^1_1(s).
\end{gather}
\end{widetext}
By a straightforward substitution, we obtain
\begin{widetext}
\begin{eqnarray}
\langle \chi_0^1(s)|\frac{d}{ds}|\chi_{0}^1(s)\rangle &=&a^{1\ast}_0(s)\frac{d}{ds}a^{1}_0(s)+b^{1\ast}_0(s)\frac{d}{ds}b^1_0(s)=0,\\
\langle \chi_1^1(s)|\frac{d}{ds}|\chi_{1}^1(s)\rangle &=&a^{1\ast}_1(s)\frac{d}{ds}a^1_1(s)+b^{1\ast}_1(s)\frac{d}{ds}b^1_1(s)=i\frac{d}{ds}\omega^1_1(s).
\end{eqnarray}
\end{widetext}
Thus the integrals are
\begin{eqnarray}
\int_0^1\langle\chi_0^1(s)|\frac{d}{ds}|\chi_{0}^1(s)\rangle
ds&=&0,\\
\int_0^1\langle\chi_1^1(s)|\frac{d}{ds}|\chi_{1}^1(s)\rangle
ds&=&i\omega^1_1(s)|_0^1=i\pi.
\end{eqnarray}
In the same manner, we calculate the contributions of the other
three line segments. The results are:
\begin{eqnarray}
\int_0^1\langle\chi_0^2(s)|\frac{d}{ds}|\chi_{0}^2(s)\rangle
ds&=&0,\\
\int_0^1\langle\chi_1^2(s)|\frac{d}{ds}|\chi_{1}^2(s)\rangle
ds&=&0,
\end{eqnarray}
\begin{eqnarray}
\int_0^1\langle\chi_0^3(s)|\frac{d}{ds}|\chi_{0}^3(s)\rangle
ds&=&i\frac{\pi}{2},\\
\int_0^1\langle\chi_1^3(s)|\frac{d}{ds}|\chi_{1}^3(s)\rangle
ds&=&0,
\end{eqnarray}
\begin{eqnarray}
\int_0^1\langle\chi_0^4(s)|\frac{d}{ds}|\chi_{0}^4(s)\rangle
ds&=&0,\\
\int_0^1\langle\chi_1^4(s)|\frac{d}{ds}|\chi_{1}^4(s)\rangle
ds&=&i\frac{\pi}{2}.
\end{eqnarray}
Putting everything together, for the diagonal elements of the
holonomy we obtain
\begin{gather}
U^{\gamma}_{0k,0k}=e^{i\frac{\pi}{2}},\notag\\
U^{\gamma}_{1k,1k}=e^{i\frac{3\pi}{2}}.\label{finalholonZ}
\end{gather}
The holonomy transforms any state in the ground space of the
initial Hamiltonian as
\begin{equation}
U^{\gamma}\sum_{jk}\alpha_{jk}|j\rangle|\widetilde{\psi}_{jk}\rangle
=
e^{i\frac{\pi}{2}}\sum_{jk}(-1)^j\alpha_{jk}|j\rangle|\widetilde{\psi}_{jk}\rangle,
\hspace{0.2cm} j=0,1.
\end{equation}
From the point of view of the full Hilbert space, this is
effectively a $Z$ gate on the first qubit up to an overall phase.

Note that other single-qubit transformations such as the Hadamard or
the $X$ gates, under which the eigenspaces of the Hamiltonian do not
follow complete loops, can be obtained in a similar fashion by
calculating the open-path expression \eqref{openpathholonomy}. In
principle, the result of that calculation depends on the choice of
basis $\{|\alpha; \lambda\rangle \}$ which is defined up to a
unitary gauge transformation. However, if the final eigenspace has a
non-zero overlap with the initial one, this ambiguity can be removed
by defining the frame in the final eigenspace to be the one which is
``most parallel'' to the initial frame \cite{KAS06}. In the case of
the single-qubit Hadamard gate, there is a non-zero overlap between
the initial and final eigenspaces, and the set of initial basis
states $|0k;0\rangle=|0\rangle|\tilde{\psi}_{0k}\rangle$,
$|1k;0\rangle=|1\rangle|\tilde{\psi}_{1k}\rangle$,
$k=1,...,2^{n-2}$, can be seen to be most parallel to the set of
final basis states
$|0k;1\rangle=|+\rangle|\tilde{\psi}_{0k}\rangle$,
$|1k;1\rangle=|-\rangle|\tilde{\psi}_{1k}\rangle$,
$k=1,...,2^{n-2}$, respectively (for a precise definition of ``most
parallel'' see Ref.~\cite{KAS06}). Thus, the resulting open-path
holonomy corresponds to flipping the phase of half of the basis
vectors. For the single-qubit $X$ gate, however, the final ground
(excited) space is orthogonal to the initial ground (excited) space.
Thus a gauge invariant expression for the geometric transformation
cannot be defined in this way.

We emphasize that the expression \eqref{openpathholonomy} for the
open-path geometric transformation taking place inside each
eigenspace of the Hamiltonian, whether gauge-invariant or not, is
not the same as the geometric transformation being realized inside
the code space which is what eventually gives rise to the logical
transformation. For example, the single-qubit $Z$ gate can be
understood as a closed-loop holonomy in the eigenspaces of the
Hamiltonian $-Z\otimes\tilde{G}$, but under this transformation the
code space does not follow a loop; in fact, it becomes orthogonal to
the initial code space so that a gauge-invariant expression for the
geometric transformation taking place inside it cannot be defined.
However, if a sequence of single-qubit $Z$ gates on all qubits
yields, say, an encoded $Z$ gate (as in the case of CSS codes), then
after we complete such a sequence we will obtain a non-trivial
closed-loop holonomy in the code space which is equal to the encoded
$Z$.

\subsection{Unitary interpolation}

The calculation is simpler if we choose a unitary interpolation,
\begin{equation}
f(t)=\cos{\frac{\pi t}{2 T}},\hspace{0.4cm} g(t)=\sin {\frac{\pi
t}{2 T}}.
\end{equation}
Such an interpolation corresponds to a rotation of the Bloch sphere
around a particular axis for each of the segments of the loop. The
first two segments of the loop \eqref{Zgate} are realized via the
Hamiltonian
\begin{equation}
\widehat{H}_{1,2}(t)=-V_{Y}^{\dagger}(t)ZV_{Y}(t)\otimes
\widetilde{G}, \hspace{0.2cm} V_{Y}(t)=\textrm{exp}
\left(it\frac{\pi }{2T}Y\right),\label{HZ1}
\end{equation}
applied for time $T$, and the third and fourth segments are
realized via the Hamiltonian
\begin{equation}
\widehat{H}_{3,4}(t)=-V_{X}(t)ZV_{X}^{\dagger}(t)\otimes
\widetilde{G}, \hspace{0.2cm} V_{X}(t)=\textrm{exp}
\left(it\frac{\pi }{2T}X\right),\label{HZ1}
\end{equation}
again applied for time $T$. Let us define the eigenstates of the
Hamiltonian along the first two segments as
\begin{equation}
|\chi^{1,2}_0(t)\rangle = V_{X}(t)|0\rangle, \hspace{0.2cm}
|\chi^{1,2}_1(t)\rangle = V_{X}(t)|1\rangle, \hspace{0.2cm}0\leq
t\leq T,
\end{equation}
and along the third and fourth segments as
\begin{equation}
|\chi^{3,4}_0(t)\rangle = -iV^{\dagger}_Y(t)Y|0\rangle,
\hspace{0.2cm} |\chi^{3,4}_1(t)\rangle =-i
V^{\dagger}_{Y}(t)Y|1\rangle, \hspace{0.2cm}0\leq t\leq T.
\end{equation}
Notice that
\begin{equation}
|\chi^{1,2}_0(T)\rangle=-iY|0\rangle=|\chi^{3,4}_0(0)\rangle,
\hspace{0.2cm}
|\chi^{1,2}_1(T)\rangle=-iY|1\rangle=|\chi^{3,4}_1(0)\rangle,
\end{equation}
but
\begin{equation}
|\chi^{1,2}_0(0)\rangle=|0\rangle \neq |\chi^{3,4}_0(T)\rangle =
-i|0\rangle, \hspace{0.2cm} |\chi^{1,2}_1(0)\rangle=|1\rangle \neq
|\chi^{3,4}_1(T)\rangle = i|1\rangle,
\end{equation}
i.e., this basis is not single-valued. To make it single valued,
we can modify it along the third and fourth segments as
\begin{equation}
|\chi^{3,4}_0(t)\rangle\rightarrow
|\widetilde{\chi}^{3,4}_0(t)\rangle =
e^{i\omega_0(t)}|\chi^{3,4}_0(t)\rangle, \hspace{0.2cm}
|\chi^{3,4}_1(t)\rangle\rightarrow
|\widetilde{\chi}^{3,4}_1(t)\rangle =
e^{i\omega_1(t)}|\chi^{3,4}_0(t)\rangle,
\end{equation}
where
\begin{equation}
\omega_0(0)=0, \hspace{0.2cm} \omega_0(T)=\frac{\pi}{2},
\end{equation}
\begin{equation}
\omega_1(0)=0, \hspace{0.2cm} \omega_1(T)=-\frac{\pi}{2}.
\end{equation}

The expression \eqref{diagonalU} then becomes
\begin{equation}
U^{\gamma}_{jk,jk}=\textrm{exp}\left(\int_0^T \langle
\chi_j^{1,2}(t)|\frac{d}{dt}|\chi_{j}^{1,2}(t)\rangle dt +
\int_0^T \langle
\chi_j^{3,4}(t)|\frac{d}{dt}|\chi_{j}^{3,4}(t)\rangle dt +
(-1)^j\frac{\pi}{2}\right), \hspace{0.2cm} j=0,1.
\end{equation}
But
\begin{equation}
\langle \chi_j^{1,2}(t)|\frac{d}{dt}|\chi_{j}^{1,2}(t)\rangle =
-i\frac{\pi }{2T}\langle j|Y|j\rangle =0,
\end{equation}
and
\begin{equation}
\langle \chi_j^{3,4}(t)|\frac{d}{dt}|\chi_{j}^{3,4}(t)\rangle =
i\frac{\pi }{2T}\langle j|YXY|j\rangle =0.
\end{equation}
Therefore, we obtain Eq.~\eqref{finalholonZ}.

\end{document}